# Effect of carbon in severe plastically deformed metals

*Andrea Bachmaier\*, Reinhard Pippan, Oliver Renk*

Dr. Andrea Bachmaier, Author 1
Prof. Reinhard Pippan, Author 2
Dr. Oliver Renk, Author 3
Erich Schmid Institute of Materials Science, Austrian Academy of Sciences
Jahnstraße 12, 8700 Leoben, Austria
E-mail: andrea.bachmaier@oeaw.ac.at

## Abstract

In the last decades severe plastic deformation techniques have gained increasing interest as they allow to produce bulk nanostructured materials with superior mechanical and functional properties. However, because of mechanically induced grain boundary migration, the achievable grain size reduction is not indefinite but tends to stagnate once sufficient strain has been applied. Consequently, addition of solute elements or second phase particles offers the possibility to access the true nanocrystalline regime. Due to their low solubility and high mobility interstitial elements are extremely effective to subdue boundary migration. Here we summarize the effect of carbon on grain refinement and the resulting mechanical properties. As carbon may not only be added as graphite but could also be introduced in other forms or allotropes such as nanotubes, nanodiamonds or carbides, the respective advantages and problems associated with it will be the center of discussion. Independent of the strategy used, strength levels hardly achievable with other alloying elements can be obtained. Moreover, as carbon does not have a negative effect on grain boundary cohesion, despite the enormous strength levels even ductility and toughness can be widely maintained.

# 1. Introduction

In the last three decades a variety of new plastic deformation techniques have been developed which permit very heavy plastic deformation. They allow for significantly larger strains compared to standard hot and cold deformation techniques like extrusion, rolling or forging, even at relatively low temperatures. These techniques are now called severe plastic deformation, SPD. It has been shown in a vast number of papers that these new techniques enable to generate ultrafine-grained (UFG) or nanocrystalline (NC) structures for most metals and alloys, usually only available in a micro crystalline bulk form (see for example [1]–[4]).

In single phase alloys and pure metals, the grain refinement process during SPD is quite similar, if dislocation based plasticity dominates [5]–[7]. In the early stages of SPD the generated dislocations arrange in cells with low misorientations between the cells and cell blocks with larger misorientations. With increasing strain, the cell size decreases and the misorientation between the cell and cell blocks increases until finally these low angle boundaries are transformed into high angle grain boundaries [5][8]–[11]. The refinement process reaches a saturation at very high strains, at low homologous temperatures typically for strains larger than 20. This saturation is established by a dynamic equilibrium between the generation and annihilation of new boundaries, dislocations and vacancies and their annihilation [8][11][12]. At low homologous SPD temperatures the annihilation processes are mainly driven by the applied stresses and strains, supported by thermal activation. Hence, the minimum grain size achievable at very high strains, when the dynamic equilibrium is reached, is determined by the thermal and the mechanical driven grain boundary migration. Due to the thermal contribution a reduction of the SPD temperature results in finer microstructures. However, below a homologous temperature of about 0.1 the thermal contribution to the local coarsening process becomes relatively small [13] and the temperature dependence of the saturation grain size eventually diminishes at these very low deformation temperatures. Independent of the dominant processes controlling grain boundary motion, strategies to subdue this migration result in a reduction of the saturation grain sizes during SPD. Beside the SPD temperature alloying or doping are very efficient ways to reduce grain boundary mobility and thus to further reduce the saturation grain size, proven in many studies for a variety of metallic materials, see for example [14]–[17]. To discuss the observed phenomena it is helpful to divide the approaches into different classes such as single phase solid solutions based on substitutional or interstitial alloying elements, multiphase or precipitation forming alloys, composites, immiscible metals etc. This enormous variety of possible starting structures makes it, except for pure metals and single phase materials, difficult to predict the resulting microstructural changes and the minimum grain size achievable during SPD.

In the present overview we will demonstrate how different types of carbon, one of the most important alloying elements, affect the structural evolution and the microstructures at very high strains. Accordingly, the description is roughly divided into three parts. First, the effect of carbon in solid solution will be analyzed. In most metals the solubility of carbon at room temperature is negligible, only at high temperatures many metals exhibit a small solubility. As an example for these types of metals carbon doped nickel will be considered here in detail. The second group to be discussed are carbon metal composites, where carbon nanotubes and nanodiamonds are introduced to stabilize or pin the grain boundaries. These examples, based on nickel and silver, can also be considered as model systems for all stable and undeformable nanocarbides. The last class which will be considered in more detail are alloys containing deformable carbides. In particular the focus will be laid on iron based systems here, although the results are generally applicable.

The paper will demonstrate on these examples how carbon affects the refinement process and the minimum grain size which can be obtained during SPD. Since the smallest structural sizes can be achieved by high pressure torsion (HPT), we focus here mainly on this SPD technique. It will be shown that carbon, like many other interstitial elements, can permit to generate significantly finer microstructures than substitutional alloying elements allow for. However, the advantage of carbon is that it often increases the grain boundary strength whereas other interstitials like hydrogen or oxygen are known to embrittle the grain boundaries.

## 2. Effect of elemental carbon on grain refinement and mechanical properties

As noted in the introductory section, for single phase materials (thermo)mechanically induced grain boundary migration dictates the minimum grain size achievable after severe strains [8][11]. Consequently, all measures to slow down boundary mobility yield also smaller grain sizes after SPD. In this chapter we will present results emphasizing the efficiency of interstitial alloying elements such as carbon, to achieve enormous grain refinement. Most structural metals have a negligible solubility for interstitials, especially at low temperatures, but SPD has been shown to increase their solubility (e.g. for carbon [18]). As interstitial alloying elements possess a much larger mobility compared to substitutional elements, they should preferentially interact with dislocations or grain boundaries. As an example to show the potential of interstitials we present here results obtained on carbon doped nickel. One batch of high purity nickel samples (99.99 %) was doped with 1200 ppm (by weight) of carbon (graphite) using repetitive arc melting. As will be shown later, no second phases such as carbides could be detected after SPD processing or remained at least below the resolution limit of the techniques used. While in case of the high purity samples HPT at room temperature (10 rotations at room temperature at an angular speed of 0.2 rot min$^{-1}$) allows for an increase of the hardness up to 3.19 GPa, the addition of only 1200 ppm carbon increases this value to 5.15 GPa when the same HPT process parameters are used. A comparison with HPT deformed nickel alloys containing substitutional alloying elements or electrodeposited NC nickel samples clearly indicates that carbon doping is a highly efficient way of reducing the grain boundary mobility, thus the achievable grain size, compare **Table 1**.

The pronounced increase in hardness for the carbon doped nickel samples is also reflected in a significant reduction of the boundary spacing, see **Figure 1 a-b**. From the inverse pole figure maps taken along the radial direction of the HPT disk, the reduction of the minimum grain dimensions from 160 nm in case of the pure nickel samples to 90 nm for the carbon doped samples is clearly visible. The reasons for this enormous effect on grain refinement can be related to the large segregation tendency of carbon to grain boundaries in nickel. Segregation energies are with about -0.8 eV much larger than for other typical substitutional alloying elements in nickel [19]. Together with the enhanced mobility of interstitial atoms already at lower temperatures this could lead to a situation where the migrating grain boundary becomes actively trapped by the carbon atoms. Similar findings were also reported for substitutional alloying elements, where a slight increase of the HPT temperature surprisingly reduced the grain size due to the enhanced segregation level [20]. In case of the carbon doped nickel samples this is supported by the atom probe tomography (APT) data gained on these samples, exemplarily shown in **Figure 1c**. As can be seen from the carbon map of the 3D reconstruction of the APT tip, carbon is mainly present at defects. Although no direct correlative investigations have been performed with the measured tips, a comparison of the box size (63 x 65 x 250 nm³, Ref. [15]) with the minimum grain dimensions measured, suggests that carbon is mainly present at the generated grain boundaries. From line profiles taken across the boundaries, local carbon concentrations of up to 3 at. % can be measured, while inside the grains only a limited amount of carbon atoms is present. From the APT analysis it is also evident that no carbides formed during the HPT process.

It should be noted that the effects of interstitial alloying elements such as carbon can outperform the effect of other ones additionally present in the material. This can lead to intriguing effects as observed in Ref.[15]. While generally a reduction of the purity level is expected to allow for further grain refinement upon SPD processing, in the mentioned work the opposite was found. On HPT deformed nickel samples with a nominal purity of 99.79 % an almost 30 % higher saturation hardness than for less pure nickel samples (99.69 %) was measured. Detailed chemical analysis revealed a six times larger carbon content in case of the less pure samples, explaining the observed effect [15][21]–[26].

As can be noted from **Table 1**, not only carbon but also other interstitial elements are efficient to reduce the saturation grain size. Using pure nickel powders as a precursor, consolidation and subsequent deformation by HPT leads to hardness levels of 5 GPa and consequently a significantly reduced grain size [22][23]. In case of the powder precursors, the native oxide layer present on the surface of the powder particles acts as an oxygen source. Although some of these oxides may fracture during processing and remain as nanoparticles at the grain boundaries, detailed studies indicate that the severe strains applied induce dissolution of the oxygen into the lattice [27]. Hence, the use of metallic powders instead of their bulk counterparts allows to make use of the effectiveness of interstitials on grain refinement and to access fundamentally different grain size levels [22][28]. However, different to oxygen or hydrogen, known to weaken the grain boundary strength for many metals, ab initio calculations provide evidence that carbon can increase the cohesive strength of grain boundaries in nickel [19]. This effect could help not only to improve strength but also to retard failure of the grain boundaries upon loading, hence to maintain ductility or toughness on a reasonable level.

To this end, mechanical tests were conducted on the carbon doped nickel samples to reveal its effect on the mechanical properties. Despite the effective strengthening induced by carbon doping, reaching almost twice the strength levels achievable for HPT deformed high purity nickel, miniaturized tensile tests conducted on both samples still indicate an excellent ductility of the doped samples (**Figure 2**). This is intriguing as strength and ductility are mutually exclusive properties. However, as can be seen from **Figure 2**, ultimate tensile stresses of the carbon doped samples reach almost 1.7 GPa, compared to slightly less than 1 GPa in case of the pure nickel samples. Despite the 70 % higher strength, the elongation to failure in case of the carbon doped samples remains at almost the same level as for the pure nickel samples and huge reduction in area of 70 % can be reached [15]. As the main difference between the two samples is the excess of carbon at the grain boundaries, the excellent mechanical performance needs to be attributed to this carbon excess. Its mentioned positive effect on grain boundary strength could help to retard damage and failure along the grain boundaries, the predominant fracture mode for nanostructured metals, as evidenced from experiments and atomistic simulations[29]–[32]. Beside the efficient strengthening while maintaining ductility, also the thermal stability of the nanostructures can be distinctively increased by carbon doping. While in case of the HPT deformed pure nickel samples already above 423 K pronounced grain growth occurs, the presence of carbon at the interfaces allows for a significantly better thermal stability, compare **Figure 3**.

The enormous efficiency of carbon doping to further reduce the saturation grain size upon SPD raises the question if these limits can be pushed even further by increasing the carbon content. Hence, a similar set of experiments was performed using nickel samples with even higher carbon concentrations. Nickel samples containing 2500, 5000 and 7500 ppm carbon were synthesized by repetitive arc melting and subsequently deformed by HPT at ambient temperature. The effect of the carbon concentration on the achievable hardness is presented in **Figure 3b**. As can be noted, hardness increases continuously with an increase of the carbon content (**Figure 3b)**. While up to carbon concentrations of 2500 ppm hardness increases almost linearly, a further increase of the carbon content becomes less effective, although further strengthening is achieved. In fact, for concentrations of 7500 ppm, hardness levels of almost 8 GPa can be reached. However, it should be noted that for these samples (7500 ppm carbon) the measured hardness may not correspond to the achievable saturation hardness. In this case the hardness of the samples reached the one of the anvils and partial slippage, limiting further deformation of the HPT disk, occurred.

As a whole, the use of interstitial alloying elements represents an effective approach to enormously reduce the achievable saturation grain size, but also to expand the thermal stability range. Nevertheless, carbon additionally offers a beneficial effect on mechanical properties due to its strengthening effect for grain boundaries, in line with the results of the tensile tests. It could thus be of interest to use carburization treatments in combination with severe surface deformation techniques such as SMAT [33], to generate wear resistant nanograined metal surfaces. Although within this chapter the effect was mostly discussed for nickel samples as a model material, the presented approach should be generally applicable. Results to be discussed in the remainder (chapter four) support this idea, as enormous grain refinement using dissolved carbon is also observed for quenched low carbon steels [34].

## 3. Carbon metal composites

SPD can be also used to synthesize nickel matrix composites reinforced with various carbon nanoparticles such as carbon nanotubes or nanodiamonds. Beside the SPD deformation temperature, the final microstructure and thus the properties of these nickel composites depend on the properties of the reinforcing phase, its amount as well as the geometry and distribution of the dispersed phase. In the following, the critical issues in HPT processing (distribution and deterioration of the reinforcement phase, interface between matrix and reinforcement phase) and the effect of the reinforcement phase on microstructure (refinement process and minimum grain size) and composite properties (mechanical properties, microstructural stability, tribological properties) are discussed.

Up to now, the most common solid state processing method for carbon nanoparticle based composites is ball milling [35]. Both processing methods, SPD and ball milling, are similar in the amount of applied strain and have the advantage that any composition can be synthesized as both processes are not dictated by phase diagrams or thermodynamic considerations. An additional benefit of SPD processing is that the dispersion of the reinforcement phase and the density of the final composites is not depending on the initial dimensions of the starting powders and no sintering treatment is necessary to achieve a bulk composite. SPD processed composite are usually also porosity free.

In general, the application of carbon nanoparticle reinforcements in composites is driven by their exceptional properties. A multi-walled carbon nanotube consists of multiple concentric graphene layers and exhibits an exceptionally high strength and elastic modulus (≈1000 GPa), a high thermal conductivity, a structural stability up to 4000 K and almost zero coefficient of thermal expansion. Multi-walled carbon nanotubes have a diameter between 4 and 30 nm and can be up to 1 μm long [35][36].

Nanodiamonds are particles with a size less than 10 nm, a small aspect ratio and a diamond-like atomic structure ($sp^3$ hybridization). They also exhibit excellent mechanical and optical properties, which can be influenced by their tunable surface structure (adsorbed atoms, molecules and functional groups)[37][38]. Carbon nanotubes and nanodiamonds can be also differentiated by their dimensionality, where the first is considered to be a 1D, the latter a 0D structure.

The matrix material of the processed composites can benefit from the superior mechanical and thermal properties of carbon nanotubes and nanodiamonds[35][39][40]. To maximize the benefit from the nanocarbon reinforcement phase, a homogeneous distribution in the nickel matrix is essential. Carbon nanotubes, however, have a tendency to form large agglomerates due to strongly attractive van der Waals forces, which deteriorate the properties of the composites[35]. Larger agglomerates are also common in nanodiamond reinforced composites due to their large surface to volume ratio[37].

In nickel matrix composites, the HPT deformation temperature has the greatest influence on the saturation grain size of the nickel matrix[41]–[43]. To obtain the nickel matrix composites, powder blends of nickel (Alfa Aesar, mesh −325, dendritic) and carbon nanotubes (CCVD grown, Graphene Supermarket, USA) with different amount of carbon nanotubes were HPT processed at different deformation temperatures. **Figure 4**, Transmission Kikuchi diffraction (TKD) orientation maps of nickel matrix composites with 2 wt% carbon nanotubes in the as-deformed state after HPT processing at 200°C, 300°C and 400°C are shown. It is evident that the nickel matrix grain size increases with increasing deformation temperature. The nickel grain sizes were determined from several TKD orientation maps for each deformation temperature as the diameter of an equivalent circle and an area-weighted lognormal distribution was fitted to the data. After deformation at 200°C, 300°C and 400°C average grain sizes of 85 nm, 293 nm and 410 nm were determined, respectively.

By contrast, the amount of carbon nanotubes did not really affect the final grain size of the nickel matrix. In Ref.[41], the evolution of the nickel matrix grain size depending on the amount of carbon nanotubes was evaluated by processing different samples with increasing carbon nanotube content with the same HPT processing parameters. No difference in the final grain size of the nickel matrix was observed. Especially for higher carbon nanotube concentrations, large carbon nanotube agglomerations persist in the composite microstructure, which have little to no influence on the grain boundary mobility and thus the final saturation grain size of the nickel matrix. The picture is different when it comes to the nanodiamonds added as a reinforcement phase. Due to their geometry (0D structure), a homogeneous dispersion of the nanodiamonds and an effective dissolution of the agglomerates in the matrix by HPT processing is more readily possible. In combination with their smaller size, a stronger influence of the amount of added nanodiamonds on the final grain size of the nickel matrix is found. **Figure 5** shows the final microstructure of nickel matrix composites reinforced with 0.5, 1, 3 and 10 vol% nanodiamonds HPT processed at room temperature. With increasing nanodiamond content, the final grain size of the nickel matrix clearly decreases.

The HPT processing temperature further influences the size of the remaining carbon nanotube agglomerates and their distribution in the matrix. In Ref.[42], the influence of different HPT processing conditions on the dissolution of carbon nanoparticle agglomerates and their anisotropy of distribution in the nickel matrix composites was studied. Nickel matrix composites with different amounts of carbon nanotubes were HPT deformed at a temperature of 200°C, 300°C and 400°C and using a two-temperature process. The two-temperature process consists of deformation at 400°C first, followed by a final deformation step at a temperature of 200°C[42]. Subsequently, the final microstructures were investigated in tangential, radial, and axial direction using back scattered electron images (**Figure 6**). After HPT deformation at 200°C, an inhomogeneous distribution of the carbon nanotube agglomerates in all three directions is visible. Larger and smaller agglomerates as well as areas with no discernible agglomerates are found. The size of the agglomerates and their variation in size is rather large. HPT deformation at 300°C results in fewer agglomerates with an overall smaller size. Some agglomerates are still elongated in shear direction. The picture is similar for processing at 400°C, but the agglomerates show a more spherical appearance. With the two temperature HPT process an effective way to achieve both, a homogeneous distribution of the carbon nanotubes and a small final grain size of the nickel matrix was found. Thus, an approach to solve the common agglomeration problem during processing of this type of composites is established. An effective dissolution of the carbon nanotube agglomerates is achieved by the initial deformation step at 400°C, whereby a fine nickel grain size is adjusted by the final deformation at 200°C. The nickel matrix grain size is similar for processing at 200°C (85 nm) and after processing at two temperatures (96 nm). Therefore, the final HPT deformation temperature determines the final grain size of the nickel matrix composites.

The properties of the processed composites, i.e. the mechanical and tribological properties, are influenced by the defect development of the carbon nanotubes during processing. Accordingly, the influence of the highly energetic HPT process on the structure of the carbon nanotubes in the nickel matrix composites was studied in Ref.[44]. Using Raman spectroscopy and analyzing the spectra (ID/IG ratio, G band width and peak shifting), the structural state and the evolution of carbon nanotube deterioration was monitored as a function of the applied strain for nickel matrix composites containing different amounts of carbon nanotubes. Both the applied high pressure (several GPa) and the shear deformation during HPT processing changed the structure of the carbon nanotubes from a $sp^2$ to a $sp^3$ hybridization (disordered carbon). The observed structural damage could be correlated to the well-known amorphization model proposed by Ferrari and Robertson[45]. Additionally, the deterioration of the carbon nanotubes is related to the amount of applied strain and further varies with the amount of carbon nanotubes in the nickel matrix composites. A highly important finding for HPT processing of nickel matrix composites reinforced with carbon nanotubes is that the structural transformation/deterioration of the carbon nanotubes is finished long before a saturation of the nickel matrix grain size and a homogeneous distribution of the carbon nanotubes is achieved. However, HPT processing for up to 20 rotations at room temperature does not result in complete amorphization of the carbon nanotubes as the G band value remains above 1580 cm$^{-1}$ [44].

Another generally important aspect of composites are the characteristics of the interface between the metal matrix and the reinforcement phase as the properties of the composite are directly related to it. An example is the mechanical behaviour of the HPT processed composites. The stronger the interface is, the higher its strength. To study the interface structure of the carbon nanotubes inside the agglomerates and between the nanotubes and the nickel matrix, in addition to Raman spectroscopy, high-resolution transmission electron microscopy (HR-TEM) investigations were performed for each HPT processing temperature and the two-temperature deformation process (**Figure 7**). No carbide formation or second phases along the interfaces are observed. For all processing conditions, the interface between the matrix and the carbon nanotube agglomerates is well-defined and shows a good bonding without apparent voids. After HPT processing at 200°C, the agglomerates consist of many multi-walled carbon nanotube fragments with random orientation (**Figure 7a-b**). After processing at 300°C and 400°C, smaller agglomerates with an elliptical shape are found in the nickel matrix (**Figure 7c-f**). The carbon nanotubes are also broken into fragments following the elliptical shape and resembling a fingerprint-like structure. The same observations for the internal structure of the agglomerates are made for the composite processed with the two-temperature process (**Figure 7g-h**). The shape of the agglomerates, however, is similar to the nickel composite processed at 200°C. Additionally, the spacing between the individual layers of the carbon nanotube fragments was measured from HR-TEM images. The values are 0.37±0.03 nm, 0.37±0.03 nm, 0.44 ±0.06 nm and 0.41±0.05 nm for HPT processing at 200°C, 300°C, 400°C and the two-temperatures, respectively.

In principle, the TEM investigations confirm the findings of the Raman spectroscopy. Although irreversible damage is introduced during HPT processing, no complete amorphization is observed. The multi walled carbon nanotube fragments inside the agglomerates show a typical interlayer spacing of about 0.4 nm despite the high amount of deformation and the elevated HPT processing temperatures. However, it is important to notice that it is not possible to synthesize nickel matrix composites reinforced with a homogeneous distribution of carbon nanotubes without deterioration of the reinforcement phase.

The mechanical properties of the nickel matrix composites were reported in reference[42]. With increasing applied strain, the microhardness generally increases in HPT deformed materials. In **Figure 8a**, the microhardness evolution as a function of the applied strain is plotted for a nickel matrix composite reinforced with 0.1, 0,25, 0,5, 1 and 2 wt% carbon nanotubes. After an initial higher rise, the rate of microhardness increase slows down until it saturates at a certain amount of strain. The saturation hardness in the nickel matrix composites is about 750 HV at an equivalent strain of 250. In **Figure 8a** it is clearly visible that the amount of added carbon nanotubes has no significant influence on the microhardness, which is consistent with the microstructural observations. In **Figure 8b**, the microhardness of the nickel matrix composites reinforced with 2 wt% carbon nanotubes, deformed at different HPT processing temperatures and the two-step HPT process, is shown. In this case, the microhardness decreases with increasing deformation temperature in accordance with the increasing nickel matrix grain size, but is highest after the two-temperature deformation process. The high hardness of the latter is related to the nanocrystalline grain size of the nickel matrix in combination with the improvement of the carbon nanotube distribution.

A different microhardness evolution is found for the nickel matrix composites, which are reinforced with nanodiamonds. In **Figure 9a**, the microhardness of nickel matrix composites reinforced with 0.5, 1, 3 and 10 vol% nanodiamonds, HPT processed at room temperature is plotted. A comparison with the microhardness of the nickel matrix composites reinforced with carbon nanotubes show that a lower saturation microhardness is reached for the nanodiamond reinforced composites. However, a stronger influence of the amount of added nanodiamonds on the microhardness is visible, because their amount is the main factor controlling the final nickel matrix grain size. A saturation microhardness of 525, 630, 706 and 804 HV is reached for 0.5, 1, 3 and 10 vol% nanodiamonds, respectively.

High strength and high ductility in structural materials are often mutually exclusive. To study the influence of carbon nanotube reinforcements, tensile tests have been conducted for nickel matrix composites with carbon nanotube contents of 0.25 to 2 wt% processed at different HPT temperatures and with the two-temperature process[42]. In **Figure 9b**, the engineering stress-strain curves are plotted for 0.25 and 2 wt% carbon nanotubes. Similar to the microhardness results, a strong influence of the HPT processing temperature on the tensile strength is visible. It was found that the main factor controlling the strength is the grain size of the nickel matrix (Hall-Petch strengthening). The nickel grain size is mainly determined by the HPT processing temperature, but a weak influence on the amount of added carbon nanotubes on the strength can also be monitored. For example, an ultimate tensile strength of 892 MPa and 1221 MPa is measured for the nickel matrix composites deformed at 400°C with 0.25 and 2 wt% carbon nanotubes, respectively. Orowan strengthening could be excluded as the TEM investigations have shown that the carbon nanotubes or the carbon nanotube agglomerates are not located inside the grains to enable dislocation looping. Further common strengthening mechanisms, i.e load transfer strengthening[46] as well as strengthening through impurities and solid solution of carbon[47], do not seem to have a significant influence on strength.

A strong influence of the amount of added carbon nanotubes is found for the ductility of the nickel matrix composites. Tensile specimens with a carbon nanotube content above 1 wt% already failed in the elastic regime. Nearly no ductility in tension and a brittle fracture behaviour was also found for HPT processing at 200°C. As can be seen in **Figure 9b**, higher HPT processing temperatures lead to a ductile behaviour during tensile testing. Scanning electron microscopy (SEM) investigations of the fracture surfaces displayed differently shaped carbon nanotube agglomerates, which lead to the conclusion that the remaining carbon nanotube agglomerations (their size, shape and concentration) in the nickel matrix composites mainly influence the type of fracture as they can act as micro defects and fracture nuclei. Smaller carbon nanotube agglomerates with a spherical shape, being more evenly distributed in the nickel matrix, could be achieved after HPT processing at 300°C and 400°C and lead to a good combination of strength and ductility. An exceptional high strength (yield strength of 1624 MPa) and a high ductility for this strength level (fracture strain of 3.1%) could be obtained for the nickel matrix composite reinforced with 0.25 wt% carbon nanotubes, HPT deformed with the two-temperature process.

To enable potential technological applications, microstructural stability during elevated temperatures must be guaranteed. In Ref.[43], the influence of carbon nanotube reinforcements on the thermal stability of nickel matrix composites was studied. Nickel matrix composites, HPT processed at a deformation temperature of 200°C and with the two-temperature deformation process, were annealed at 500°C (0.45 $T_m$) for 3h and their microstructural stability was compared. The amount of added carbon nanotubes ranged from 0.1 to 3 wt%. As already discussed above, a similar NC nickel matrix grain size of 85 nm (HPT deformation temperature of 200°C) and 96 nm (after HPT processing at two temperatures) could be reached. Secondly, the distribution of the carbon nanotube agglomerates on the micro- and nano-scale significantly differs between both processing conditions. However, significant grain growth and transformation to an ultra-fine grained structure (grain size above 300 nm) was measured after annealing for all carbon nanotube concentrations and both HPT processing conditions. In **Figure 10**, the microhardness as a function of the carbon nanotube content is plotted in the as-deformed and annealed state of the nickel matrix composites.

As the microhardness is extremely sensitive to microstructural changes, an enormous decrease is observed for all nickel matrix composites. Even for high carbon nanotube contents, a microhardness decrease of about 50% is monitored. To take advantage of kinetic microstructural stabilization by second phase particles during annealing, a small particle size, a high particle volume fraction and a homogeneous distribution of the particles in the metal matrix is needed[48][49]. Although a better distribution of the carbon nanotubes in the nickel matrix at the macroscale is achieved with the two-temperature process, they still show a tendency to form agglomerates at the nanoscale, which are larger than the nickel matrix grain size (**Figure 7**)[42][43]. This also explains why there is no significant influence on the amount of added carbon nanotubes on the thermal stability of the nickel matrix. As a consequence thereof, an increasing carbon nanotube content does not mean that there are more small carbon nanotube agglomerates/particles available at the grain boundaries. Most of the carbon nanotubes can be found in large agglomerates, with a size too large to account for an effective particle pinning (**Figure 7**).

Using silver as another matrix metal, nanocrystalline silver matrix composites with a homogeneous distribution (at the macro- and nanoscale) of different amounts of nanodiamonds have also been prepared by HPT[50]. In **Figure 11**, a TEM image of a silver matrix composite with 2 wt% nanodiamonds is shown, with the nanodiamonds indicated by the red lines. In this case, the microhardness only decreases by 7 and 11% after annealing at 400°C (0.54 $T_m$) and by 19% after annealing at even higher homologous temperatures (0.71 $T_m$, 600°C). For all nanodiamond reinforced silver matrix composites, NC grain sizes could be maintained after the annealing treatments. Due to the smaller size in relation to the silver matrix grain size and the uniform distribution of the nanodiamonds in the silver matrix, efficient particle pinning is observed, leading to an enhanced thermal stability.

To investigate the tribological behaviour of the nickel matrix composites, micro-tribological testing was performed using a ball on disk tribo-meter under linear reciprocating motion[51]. In this study, nickel matrix composites with 0.5, 1 and 2 wt% of carbon nanotubes, HPT processed at room temperature, were investigated. The applied strain during HPT was high enough to ensure saturation of the nickel matrix grain size. However, the steady state coefficient of friction of the nickel composites (about 0.3) did not significantly deviate from pure HPT processed nickel samples tested under the same conditions. There are two main reasons for this behaviour. First, a continuous formation of a stable oxide layer was observed. The oxide layer formation is promoted by the large grain boundary area in the HPT processed nickel and nickel composites and was experimentally confirmed by SEM investigations. Second, it is not possible to process nickel matrix composites by SPD without deterioration of the carbon nanotubes. Damaged carbon nanotubes further have a higher tendency for oxidation. Summing up, the added carbon nanotubes in the nickel composites are not suitable as solid lubricants and have nearly no influence on the steady state coefficient of friction. However, as already noted above, the HPT processed nickel composites possess high strength in combination with a high ductility. Thus, reduced wear without huge microstructural changes after tribological testing was observed.

Finally it has to be mentioned that the described phenomena are not limited to nickel and silver matrixes, but a similar behaviour can be expected for other metal and alloy matrixes. As long as the spacing of the stabilizing particles or phases during SPD is large compared to the characteristic lengths of the evolving dislocation or grain structure the microstructural evolution is similar to that of the pure metal or matrix alloy. Only when the spacing of the particles becomes smaller than these characteristic length scales, the structural evolution and the thermal stability is significantly affected by the particles. This fundamental aspect is also observed for composite systems of immiscible metals or dual phase alloys, see for example Ref.[17].

**4. Deformable and dissolvable carbides**

Carbides are frequently used in alloys as microstructural design elements to tailor the microstructure during heat treatments and processing. They are further used as strengthening elements and to improve the wear resistance of metallic materials. Usually the strength of the carbides is much larger than that of the metallic matrix. Especially for alloys consisting of a soft matrix, during plastic deformation only the matrix deforms plastically while the carbides remain undeformed very often. In such cases, even during SPD the carbides remain widely unaffected, although SPD can be used to improve the homogeneity of the carbide distribution [52][53], as described for the nanodiamonds above. The associated problems in this case are also very similar. As long as the carbides are embedded in the metallic matrix, a homogeneous distribution can be reached rapidly [53]. However, when large and dense carbide clusters are present, very high strains are required to dissolve them. The dissolution of such dense clusters usually takes place by particle erosion during the deformation process, similar as discussed for the nanodiamonds. However, carbides can be fragmented or plastically deformed too. In this chapter we will focus predominantly on such deformable carbides. Beside the strength of the matrix alloy the size and shape of the carbides are essential factors to ensure deformation of the carbides.

The technically most relevant deformable carbide in steels is cementite in different forms. Ideal co-deformation can occur in fully pearlitic steels. Similarly, the cementite in bainitic steels and the very fine cementite in martensitic steels is deformable during SPD[54]–[58]. The most important characteristic of these carbides is their plate or needle-like shape which permits to transfer sufficient stress to the high strength cementite[59]–[62]. These steel grades are very widely used in mechanical and civil engineering. In many applications they are subjected to heavily repeated contact or sliding loads (rails, wheels, roller bearings etc.). Due to this loading, the near surface layer of the contact area is subjected to very heavy plastic deformation[59][63][64]. Hence the SPD behaviour of these types of steels is not only of interest for the generation of new high strength steels but also to understand the in-service behaviour and the associated microstructural changes of the engineering components. This is important as the behaviour of the contact areas is often the lifetime controlling factor[55][59]. This industrial problem has thus induced additional attention to SPD of these types of steels.

Despite the different initial microstructures and mechanical properties of pearlitic, bainitic and martensitic steels their deformation behavior during monotonic SPD at high plastic strains is quite similar. The evolution of hardness as a function of the equivalent strain applied by HPT for these types of steels is presented in **Figure 12**[54]–[56][61].

Contrary to single phase materials, exhibiting a saturation of hardness as a consequence of the development of a dynamic equilibrium between generation and an annihilation of grain boundaries, for the pearlitic and bainitic steels the hardness continuously increases without a sign of saturation. HPT experiments are usually performed with anvils of tool steels having a hardness between 700 and 900 HV. Hence, HPT deformation is only possible as long as the strength of the sample is smaller than the strength of the tool steels. In a few selected experiments using hard metal anvils it could be shown that this hardening continues at even higher strains. These results are in line with those obtained on heavily cold drawn fully pearlitic wires[60]–[63][65]–[69], which deliver a plastically deformable material with the currently highest strength of about 6.7 GPa in tension.

The described continuous hardening is easily explainable by the structural evolution of the fully pearlitic steels as a function of the applied strain, presented in **Figure 13** and **14**. In the early stages of plastic deformation the initially randomly oriented pearlite colonies become aligned along the shear direction during HPT deformation. Only those colonies which are unfavorably oriented with respect to the global deformation fold and align along the shear direction at somewhat larger strains. This alignment continues, accompanied by a decrease of the ferrite and cementite lamellae spacing, compare Refs.[54]–[56][70].

The very strong hardening during the early stages of deformation can be explained by a simple composite model[71]–[73]. The ferrite matrix deforms plastically while the significantly stronger cementite deforms elastically, stabilizing the plastic flow of the ferrite and inducing strong hardening. Already after small degrees of plastic deformation of the softer ferrite phase even the cementite deforms plastically and the hardening is significantly reduced. Further hardening is predominantly caused by the hardening of the ferrite matrix while the contribution to the strength from the cementite remains constant. At larger strains the lamellar spacing decreases constantly, being the main reason for the continuous hardening up to very high strains. It should be noted that not only the spacing of the ferrite lamellae, but also the thickness of the cementite lamellae decreases continuously, although the detailed deformation mechanisms of the cementite remain currently poorly understood.

X-ray diffraction (XRD) and analyses of the electron energy-loss near-edge-fine structure (EEL) spectra, indicate that in the early stage of deformation the crystal structure of the cementite structure remains unchanged, for the more details of the spectra see Ref.[74]. However, at higher strains the XRD peaks of the cementite disappear as illustrated in **Figure 15.** Further, the EEL spectra indicate that the typical fingerprint of the cementite vanishes as shown in **Figure 16**[74][75]. At a strain of two the fingerprint, i.e. the values for the Fe-L2,3 peak ratio of the cementite, in the EEL spectra is clearly visible. However, at a strain of eight no Fe-L2,3 peak ratio of the cementite, or thus more correctly the carbon rich lamellae, can be found. The change of the crystal structure of the cementite is linked to a partial dissolution of the carbon into the ferrite[70][74][76]–[79]. This structural evolution during HPT is similar to that of cold-drawn pearlitic wires[67][68][80], ball-milling of pearlitic powders[81][82] or cold rolling of pearlitic steels[83][84]. It is generally assumed that the dissolution of carbon is a result of plastic deformation, however, the details of the mechanisms behind are still controversially discussed. The interested reader is referred to[65][85][86]. It should be mentioned that the necessary equivalent von Mises strain to obtain a certain lamella spacing differs significantly between rolling, wire drawing, and pure shear dominated SPD (HPT or ECAP) processes. Nevertheless, the amount of dissolved carbon is comparable for a similar lamellar spacing, independent of the deformation process. For a quantitative estimation of the lamellar spacing as a function of strain for the different deformation processes in the case of co-deformation, which is nearly fulfilled for fully pearlitic steels see[87].

For the pearlite the initial lamellae spacing is relatively unique but the colonies are randomly orientated. Therefore, the reduction of the lamellar spacing in the different colonies during deformation differs. This results in a certain distribution of the lamellar spacing[62][73], independent of the deformation process.

As mentioned the thickness reduction of the ferrite and cementite (carbon rich lamellae) does not reach a saturation during HPT deformation. With standard HPT anvils made of tool steels a ferrite lamellar spacing between 10 and 20 nm can be obtained, resulting in a similar strength than the anvil material. However, wire drawing permits to refine the structure further, and recently lamellar spacings between 5 and 10 nm could be reached[65][68]. Despite some indications that the lamellar arrangement of the ferrite and the carbon rich layer seems to transform to a more globular structure, the carbon rich layers partly neck at the highest drawing strains. The appearance of the fracture surface[88] clearly indicates that a lamellar structure still remains for 5 to 10 nm ferrite spacing. It should be noted that the thickness of the ferrite grains in this heavy deformed state is equal to the lamellar spacing. The length and width of the platelet like ferrite grains are more difficult to determine, and are thus not well documented, for details see[80][89]. However, the crystal structure of the ferrite remains unchanged, only the crystallite size becomes nanocrystalline and a supersaturation with carbon takes place. Nevertheless the carbon rich layer and segregation of carbon to the generated grain boundaries are an essential requirement to stabilize the nanocrystalline ferrite structure[57][76][80][90][91]. The importance of carbon segregation to the grain boundaries and their stabilisation effect are also observed for the HPT deformed carbon doped nickel samples presented and even more clearly visible in the case of the martensitic steel discussed in the following.

As shown in **Figure 12** also the bainitic steel, despite the different initial structure, exhibits the same hardening and a similar structural evolution at larger strains. The tiny cementite platelets become aligned along the shear direction during HPT deformation. The cementite platelets are also deformed and a lamellar structure similar to the pearlitic steels is obtained, with the only difference being a reduced length of the carbon rich layers between the ferrite lamellae[59]. But not only bainitic steels behave similarly as the fully pearlitic steel, also martensitic steels form similar structures and hardening behaviour at higher strains[34]. **Figure 17** shows the structural evolution of a quenched 0.1 wt% C martensitic steel during HPT deformation. The initial martensitic lath width was about 200 nm and the orientation of the laths was random. As clearly visible from the TEM micrographs, the lath structure has been aligned along the shear direction and the spacing of this newly formed nanolamellar structure decreases in the same way as observed for the pearlite. In the initial as-quenched martensitic state not the entire carbon content is dissolved in the matrix. A significant amount of carbon is segregated to the lath boundaries and can even form nano-platelets at these boundaries[34][92]. The segregated carbon and the nano-carbides, which may have been deformed, stabilize the nanolamellar structure in a similar way as the carbon rich layers between the carbon supersaturated ferrite lamellar do in case of deformed pearlitic or bainitic steels.

With increasing carbon content the strength of martensitic steels increases and SPD deformation becomes more and more difficult, similar as for bainitic and pearlitic steels at larger strains [92][93]. Nevertheless, the structural evolution during heavy plastic deformation remains similar. An example is depicted in **Figure 18**, taken from[92]. A commercial through-hardened 100Cr6 (AISI 52100) has been HPT deformed at a pressure of 9.5 GPa for 0.75 revolutions. The initial microstructure consisted of a martensitic matrix and about 4% of spheroidized cementite. It is evident that even this very hard martensitic matrix deforms in a similar way as the comparably soft 0.1% C martensite described above. From the micrograph in **Figure 18** it seems that the spheroidized carbides remain unaffected, they only flow in the matrix, as observed during SPD of Al metal matrix composites or in case of the nanodiamonds discussed before. However, a detailed analysis of the size distribution of the carbides indicates that their size shrinks significantly with increasing strain. The refinement of the carbides is caused by partial deformation and an erosion of the developed surface irregularities. The reason for the different behavior of these spheroidized carbides compared to other hard particles, e.g. in metal matrix composites, MMCs[53], is the significantly stronger matrix material, in other words the significantly smaller difference in strength between the matrix and the particles. This assumption is supported by Ref.[94] where carbids in a softer ferritic matrix remain widely unaffected.

Unfortunately, the conditions under which the hard carbides only flow in the metallic matrix, when they are eroded or fractured, when they plastically deform in a similar way as the surrounding matrix, or when and how they are dissolved, cannot be answered generally. However, at least for the cementite it can be summarized that small plates or lamellae can be severely deformed to very high degrees of plastic strain. The resulting lamellar structures consisting of a carbon rich interlayer, allow to stabilize lamellae with sub-10 nm spacing. Carbon rich boundaries present in martensite can stabilize the nanolamellar structure in a similar way. But one question remains still open: What is the necessary carbon content to stabilize a certain lamellar spacing?

Ductility and fracture toughness as a function of strength of pearlitic, bainitic and martensitic steels follow the typical banana-shaped curve, well-known from other materials.

However, the properties of the described nanolamellar structured steels are an exception to the strength-ductility or toughness tradeoff, clearly visible in the Ashby plot of **Figure 19**.

An important feature to realize such exceptional property combinations is the development of an anisotropy in fracture toughness and ductility for this type of materials, while the anisotropy in strength is somewhat less pronounced. **Figure 20** illustrates the development of the anisotropy in fracture toughness during HPT deformation of a fully pearlitic steel R260. If the pre-crack in the fracture toughness sample is parallel to the shear direction the fracture toughness decreases with increasing strain or increasing strength (compare **Figure 12**). For this loading direction the reduction in toughness is even somewhat more pronounced than for most other alloy classes. However, when the pre-crack is introduced perpendicular to the shear direction the fracture toughness increases initially with increasing strain, i.e. it increases with increasing strength. Only for very high strains the fracture toughness decreases somewhat, but still remains at a higher level compared to the undeformed starting material[86]. The reason for this behaviour can be rationalized from the fracture surfaces of the different sample orientations. When the pre-crack is introduced parallel to the shear plane, the crack propagates by a cleaving process along the well aligned cementite lamellae, or the carbon rich interlayer between the ferrite lamellae[95], see **Figure 21**. These results indicate that the fracture resistance in this brittle direction deceases with decreasing lamellae spacing, i.e. the work for the generation of a fracture surface decreases with deceasing lamellae spacing. The same fracture behavior is also observed in the cold drawn wires. **Figure 21c** shows the fracture surface of an experiment to determine the fracture toughness in the splitting direction (crack propagation along the wire axis) of the strongest cold drawn wire with a strength of 6.7 GPa[88]. Despite the microstructural examination exhibits some deviations from the pure lamellar structure[68] at these very high drawing strains, the fractographs indicate that the lamellar structure persists even for a lamellae spacing below 10 nm, as already mentioned. The different appearance of the fracture surface in **Figure 21a** and **c** reflects the difference in the evolution of the original pearlite colonies. In the HPT deformed pearlite a nearly ideal plane lamellar structure is formed whereas in the wire a wavy (curling) structure in thickness direction and very elongated perfectly aligned lamella structure in the axial direction has been developed. This wavy structure of the wires results in a somewhat larger fracture toughness in the splitting direction compared to the HPT deformed pearlite for the same lamellae spacing[88]. For the loading direction parallel to the shear direction of a HPT sample, i.e. a crack perpendicular to the shear plane, one has to distinguish between two extreme cases: (i) when the crack front of the pre-crack is parallel to the shear plane and (ii) when it is perpendicular to the shear plane. In both cases the fracture toughness is significantly larger than in the above described splitting direction (brittle direction). When the crack front of the pre-crack is parallel to the shear plane usually a crack deflection into the plane of low fracture toughness takes place. A crack extension in the desired direction is impossible, which does not permit to measure a valid fracture toughness value for this crack propagation direction[96]. For an orientation where both, the plane of the pre-crack and the crack front is perpendicular to the shear plane an exceptionally ductile behavior is observed. The high fracture toughness is a consequence of the delamination in the shear plane, which significantly reduces the triaxiality of the stress state ahead of the crack tip[96]. Similar delamination also takes place in the cold drawn wires as shown in **Figure 21d**. For the same lamellar spacing the fracture toughness of the cold drawn wires and the HPT deformed pearlite is similar. This delamination toughening is the main reason that this type of steels are the currently strongest damage tolerant materials.

This high fracture toughness is responsible for the relative good ductility and flaw insensitivity in the loading direction parallel to the aligned lamellar structure. For medium HPT strains the strain to uniform elongation is only moderately reduced and the reduction in area can be even increased. At higher strains the uniform elongation shrinks to very small values, which is typical for cold worked high strength materials. However, the reduction in area remains relatively high or increases sometimes, which permit relatively large plastic bending angles. These phenomena have also been observed for the cold drawn pearlitic wires[68]. It is important to note that for this behaviour the aligned lamellar arrangement is the most important feature of the heavily deformed state, while the initial microstructures before the SPD deformation are of minor relevance (compare **Figure 22a** and **b**)[59][86]. For tensile loading perpendicular to the aligned lamellar structure the situation is different, in this case a significant reduction of ductility is observed[59], which is a consequence of the above described significant reduction of fracture toughness for this loading direction.

The developed lamellar structure also induces an anisotropy of strength and flow stress in different loading directions[73][75]. Such behaviour is not surprising for a composite material with lamellar arrangement of strong and soft phases. Even an undeformed single pearlite colony exhibits a significant anisotropy[73], which only disappears in pearlitic steels due to the random arrangement of the colonies. This anisotropy however remains or even increases at higher applied HPT strains as shown in **Figure 23**, despite the significant increase of the strength of the ferrite matrix (lamellae). The main characteristics of the anisotropy can be summarized as follows. The flow stresses perpendicular and parallel to the aligned lamellae are relatively similar despite the difference in the plastic deformation pattern. If the lamellar arrangement is inclined to the loading direction the flow stress is significantly reduced. For small lamellar spacings, i.e. high pre-strains, this anisotropy can be associated with the change of the mean free path and the extension of the dislocations[73].

The strength as a function of the micro and atomistic structure of such heavily cold worked carbon steels is intensively discussed in the literature for the cold drawn pearlitic wires, see for example Refs.[61][62][65][68][72][86][89]. It is not the aim of this paper to discuss the strengthening effects in this SPD generated nanolamellar steels as the mechanisms are more or less the same as for the cold drawn wires. The contribution from the composite effect (load partitioning between ferrite and cementite) is less important or disappears at very high strains. However, the usually assumed important contributions of dislocation hardening, the effect of the dissolved carbon and the effect of the lamellar spacing (i.e. Hall-Patch hardening) are not straight forward to separate. With this overview article we do not want to contribute to this heavily discussed and unambiguously important question of the mentioned strengthening mechanisms. Nevertheless, one question should be placed. Can the strength be properly described by simply summing up the contributions of mechanisms investigated for coarse grained metals or should one rather think of how these mechanisms are subject to change once such lamellae approach the nanometer scale? This relates to the sometimes postulated idea of a transition from Hall-Petch to Orowan dominated hardening being a better approximation for such type of materials having a sub-10 nm lamellae spacing and carbon stabilized interfaces[73] [71]. Regarding this controversy the interested reader is referred to Refs.[61][62][65][68][72][86][89] and the huge literature on the strength of nano lamellar structures see for example the overview of A. Misra et al.[97][98].

## 5. Concluding remarks

In this work the effect of various types of carbon additions to metallic matrices subjected to SPD have been summarized. As other interstitial elements carbon is extremely efficient to reduce the saturation grain size. Significant effects are already noticed for contents of several 100 ppm. Also other carbon sources such as carbon nanotubes, nanodiamonds or carbides, given that dense clusters can be avoided or they co-deform, can provide effective obstacles to grain boundary motion, hence reducing the achievable grain size efficiently. This allows one to access unprecedented strength levels for the given material class. However, different than conventional strengthening approaches, the presented strategies allow to unite these high strength levels with an exceptional ductility and toughness. These property combinations can be related to the grain boundary strengthening effect of carbon and the peculiar nanolamellar structures obtained for severely deformed carbon steels, inducing an effective toughening effect. The presented examples show the enormous effect of carbon decoration of grain boundaries on the stability against mechanical driven grain boundary migration during SPD. However, for the temperature stability of the SPD generated nanostructures, carbon nanotubes or nanodiamonds are more efficient.

**Acknowledgements**

This project has received funding from the European Research Council (ERC) under the European Union's Horizon 2020 research and innovation programme (Grant No. 757333). A. Bachmaier gratefully acknowledge the financial support by the Austrian Science Fund (FWF): I2294 N36.

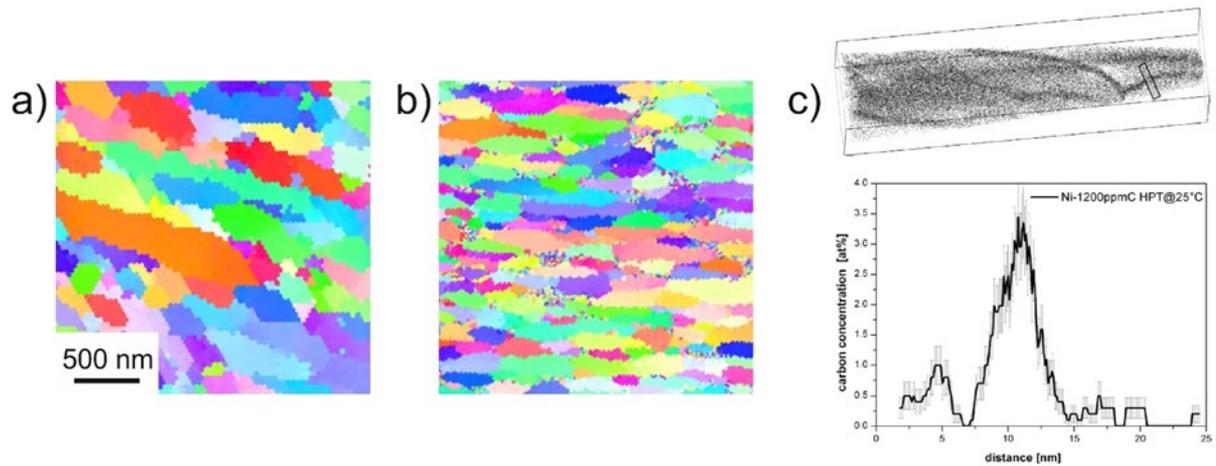

**Figure 1.** Colour IPF maps of (a) high purity nickel samples and (b) carbon doped nickel samples (1200 ppm C) subjected to HPT at RT for 10 rotations. Both images are recorded in radial direction of the HPT disk. (c) Carbon map of a reconstructed APT tip of the carbon doped sample and a representative line profile taken along one boundary. Reproduced under the terms of CC BY-NC-ND 3.0 license [15] . 2011, Elsevier.

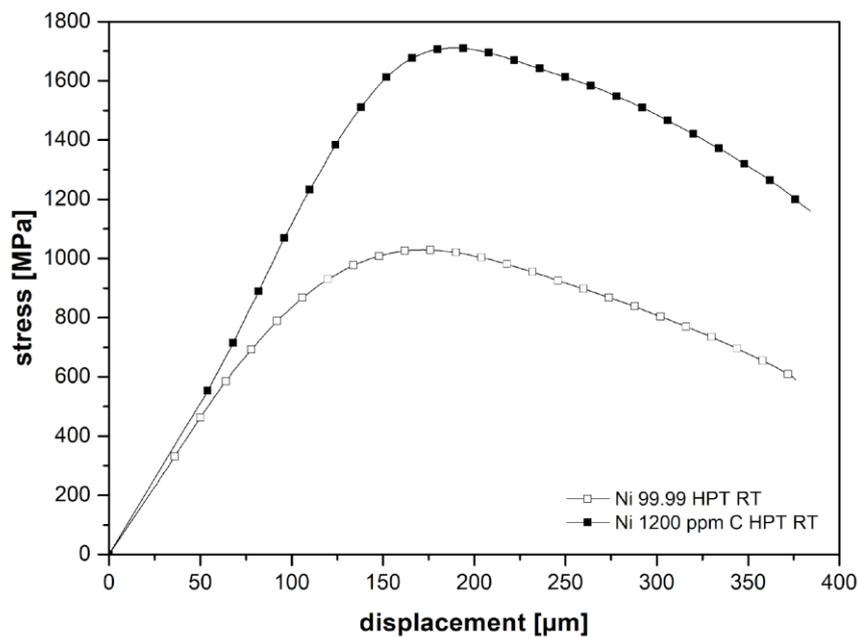

**Figure 2.** Representative engineering stress data of miniaturized tensile tests on HPT deformed pure and carbon doped nickel samples. Adapted under the terms of CC BY-NC-ND 3.0 license [15]. 2011, Elsevier.

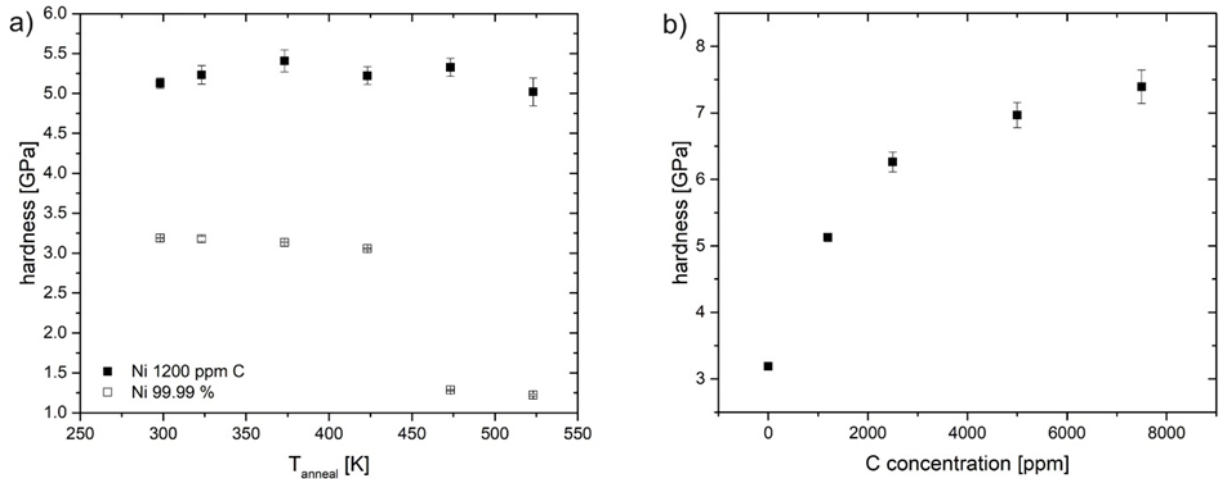

**Figure 3.** a) Comparison of the thermal stability of carbon doped (1200 ppm C) and pure (99.99 %) nickel samples deformed by HPT at room temperature using isochronal heat treatments (30 min); b) Effect of the carbon concentration on the achievable hardness after HPT at room temperature.

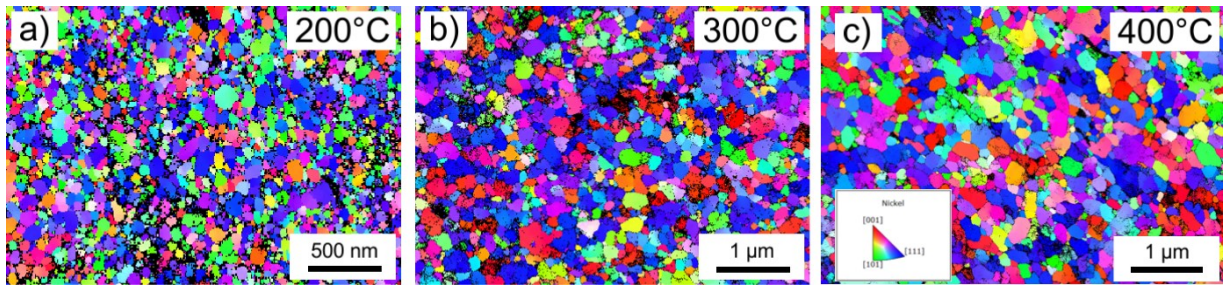

**Figure 4.** Transmission Kikuchi diffraction orientation maps of nickel matrix composites with 2wt% carbon nanotubes in as-deformed state after HPT processing at 200°C (a), 300°C (b) and 400°C (c).

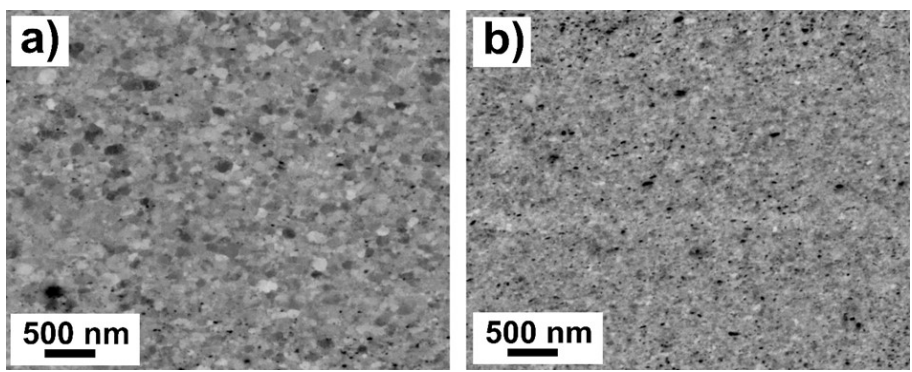

**Figure 5.** Back scattered electron images of the nickel matrix composites with 0.5 (a) and 3 vol% (b) nanodiamonds in as-deformed state after HPT processing at room temperature.

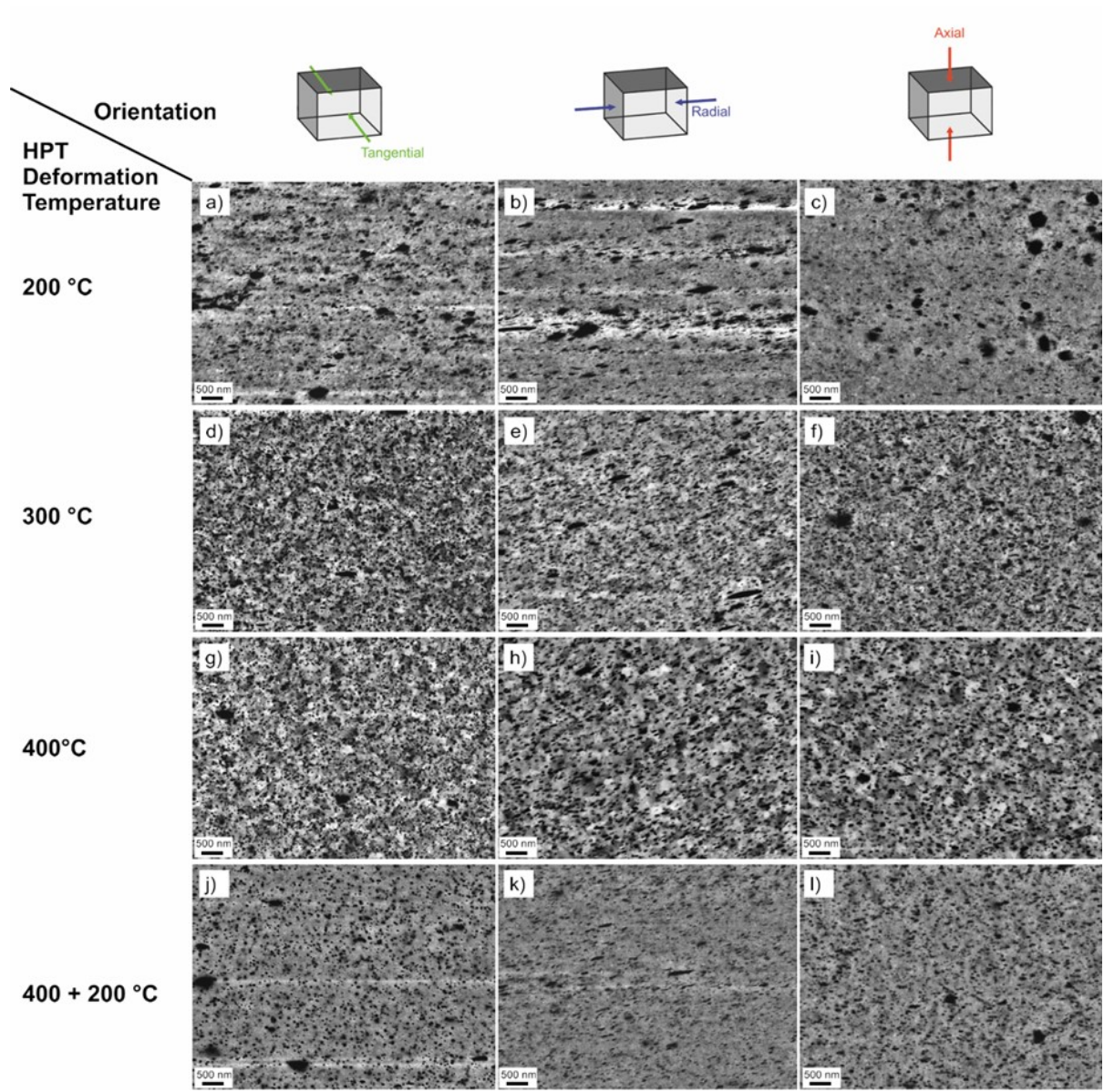

**Figure 6.** Back scattered electron images nickel matrix composites with 2 wt% carbon nanotubes in tangential, radial and axial orientation. The dark regions correspond to carbon nanotube agglomerates. The nickel matrix composites were deformed at 200°C, 300°C and 400°C and with a two-temperature process. Reproduced with permission[42]. 2018, WILEY-VCH Verlag GmbH & Co. KGaA.

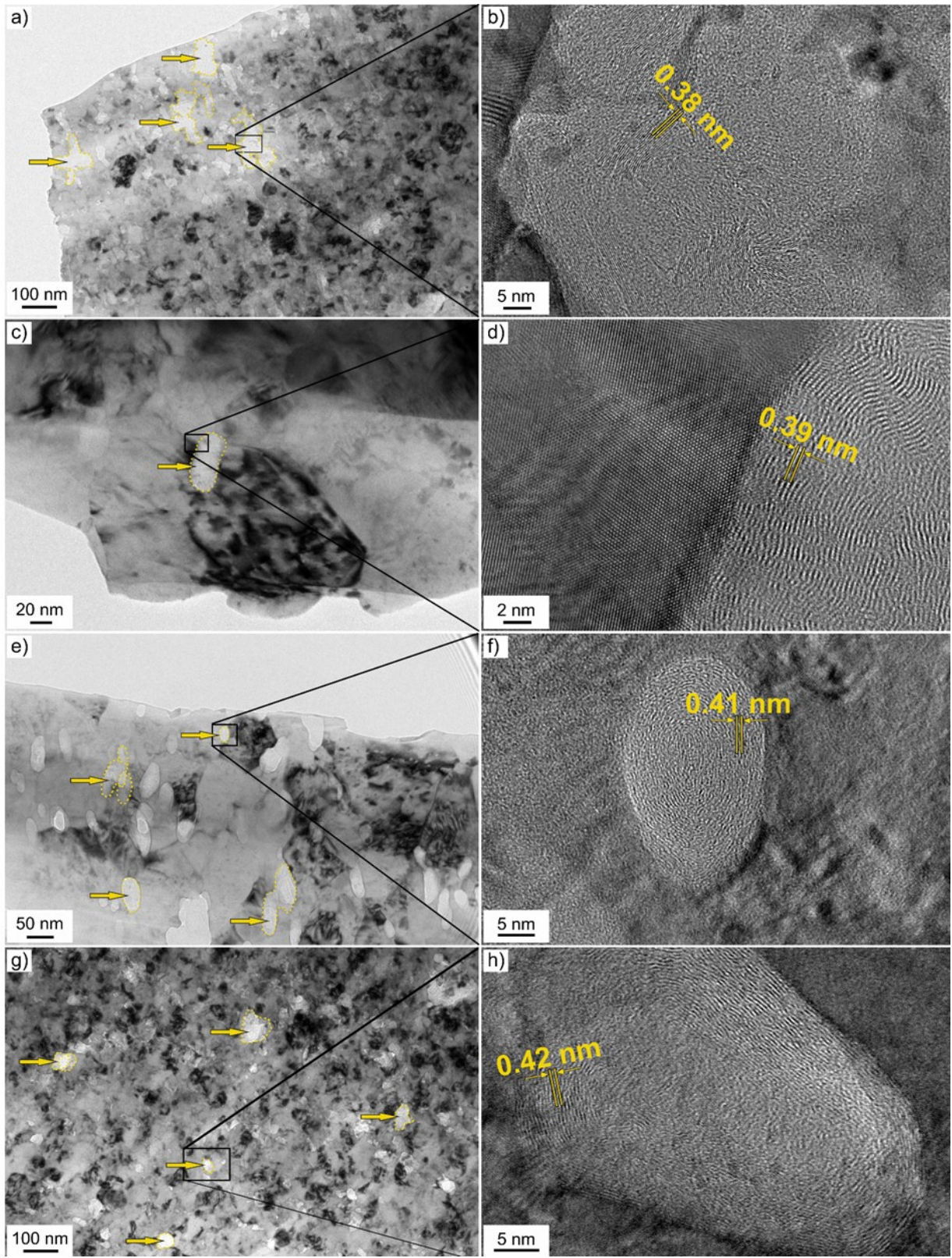

**Figure 7.** Transmission electron microscopy and high-resolution transmission electron microscopy images of nickel matrix composites reinforced with 2wt% carbon nanotubes HPT processed at 200°C (a,b), 300°C (c,d), 400°C (e,f) and the two-deformation process (g,h). Carbon nanotube agglomerates are indicated by arrows on the left side. Exemplarily measured carbon nanotube spacing are indicated on the right side.

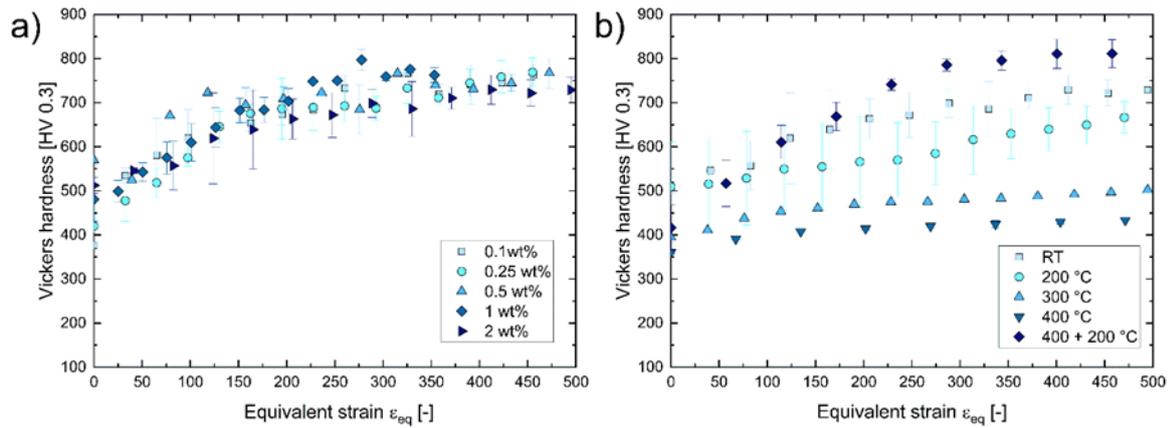

**Figure 8.** Microhardness of nickel matrix composites as a function of increasing carbon nanotube content (a) and different HPT processing temperatures (b).

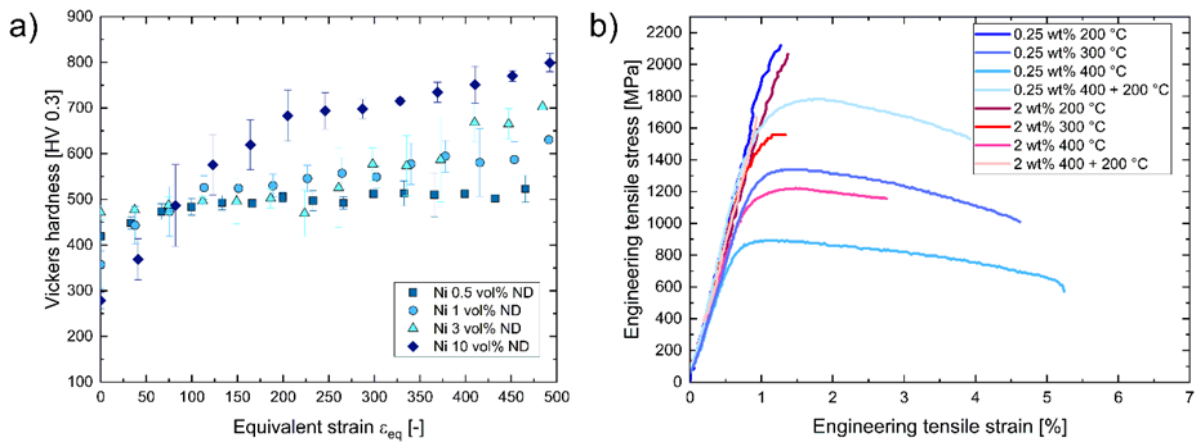

**Figure 9.** Microhardness of nickel matrix composites as a function of increasing nanodiamond content HPT processed at room temperature (a). Engineering stress-strain curves of nickel matrix composites with 0.25 and 2 wt% carbon nanotubes deformed at different HPT deformation temperatures.

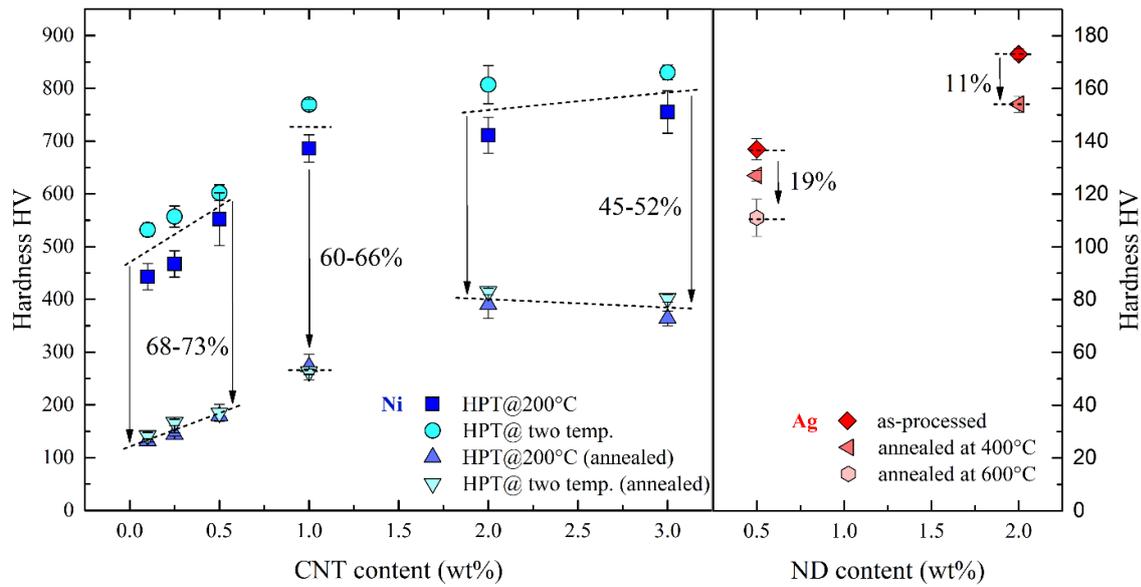

**Figure 10.** Microhardness of nickel matrix composites as a function of increasing carbon nanotube content HPT processed at 200°C and with the two-temperature process in the as-deformed and annealed state (500°C for 3h). Microhardness of silver matrix composites reinforced with 0.5 and 2 wt% nanodiamonds annealed at 400°C and 600°C. Reproduced with permission[43]. 2020, Elsevier.

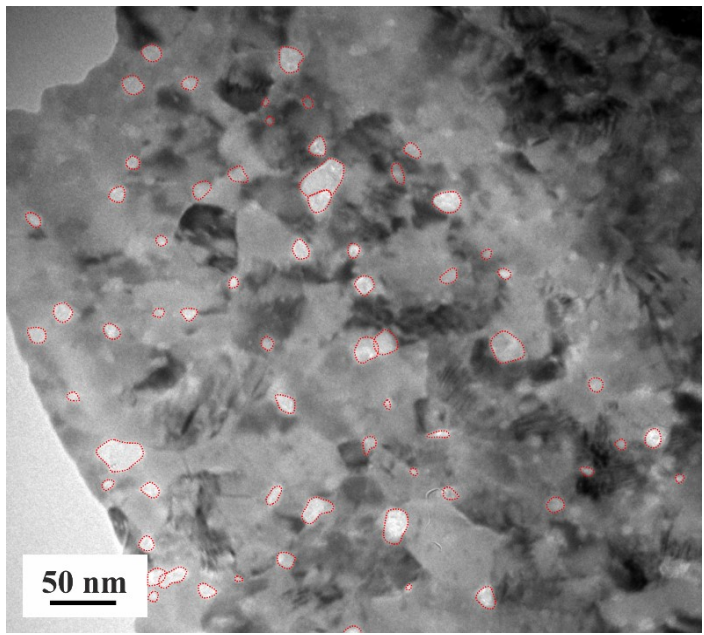

**Figure 11.** Transmission electron microscopy image of a silver matrix composite reinforced with 2 wt% nanodiamonds. Reproduced with permission[43]. 2020, Elsevier.

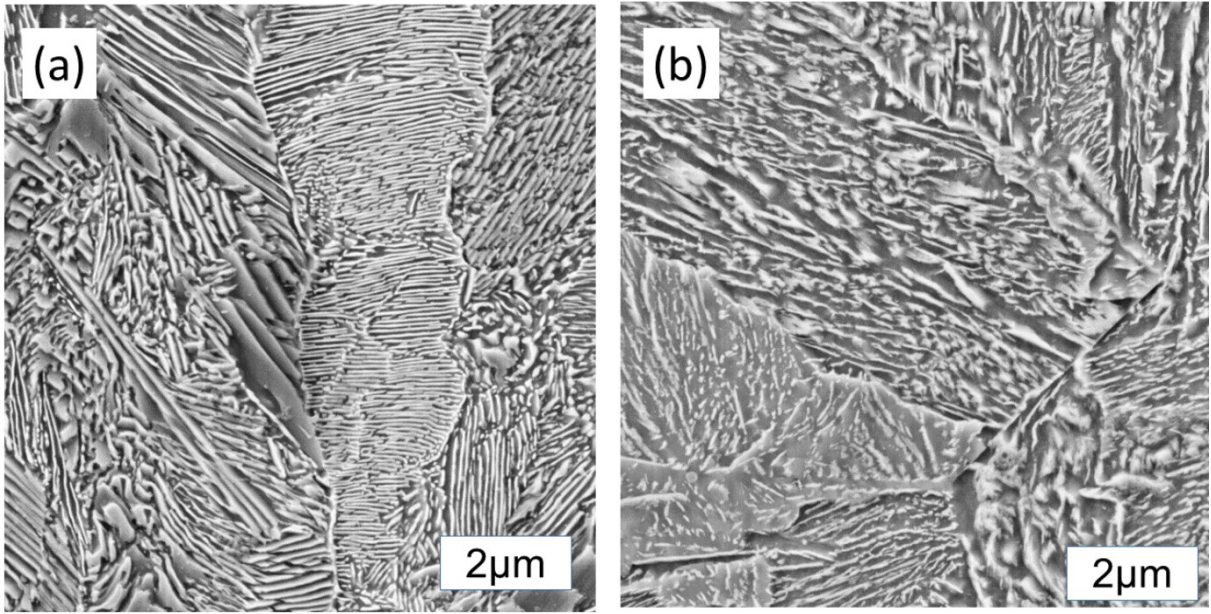
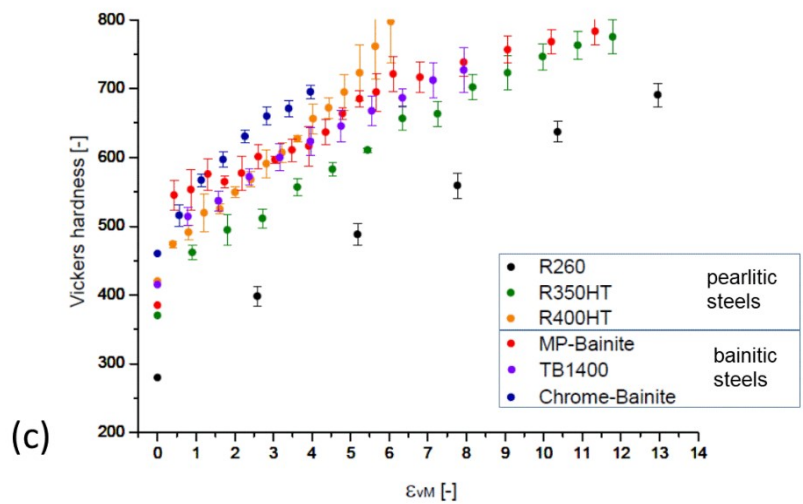

**Figure 12.** Illustration of hardening behaviour of rail steels during HPT deformation. (a) and (b) shows scanning electron microscopy micrographs of such typical pearlitic and bainitic steels, (a) R350 and (b) TB1400. (c) shows the increase of hardness as a function of the equivalent strain of different type of full pearlitic and bainitic rail steels. Reproduced with permission[59]. 2013, Kammerhofer Christoph.

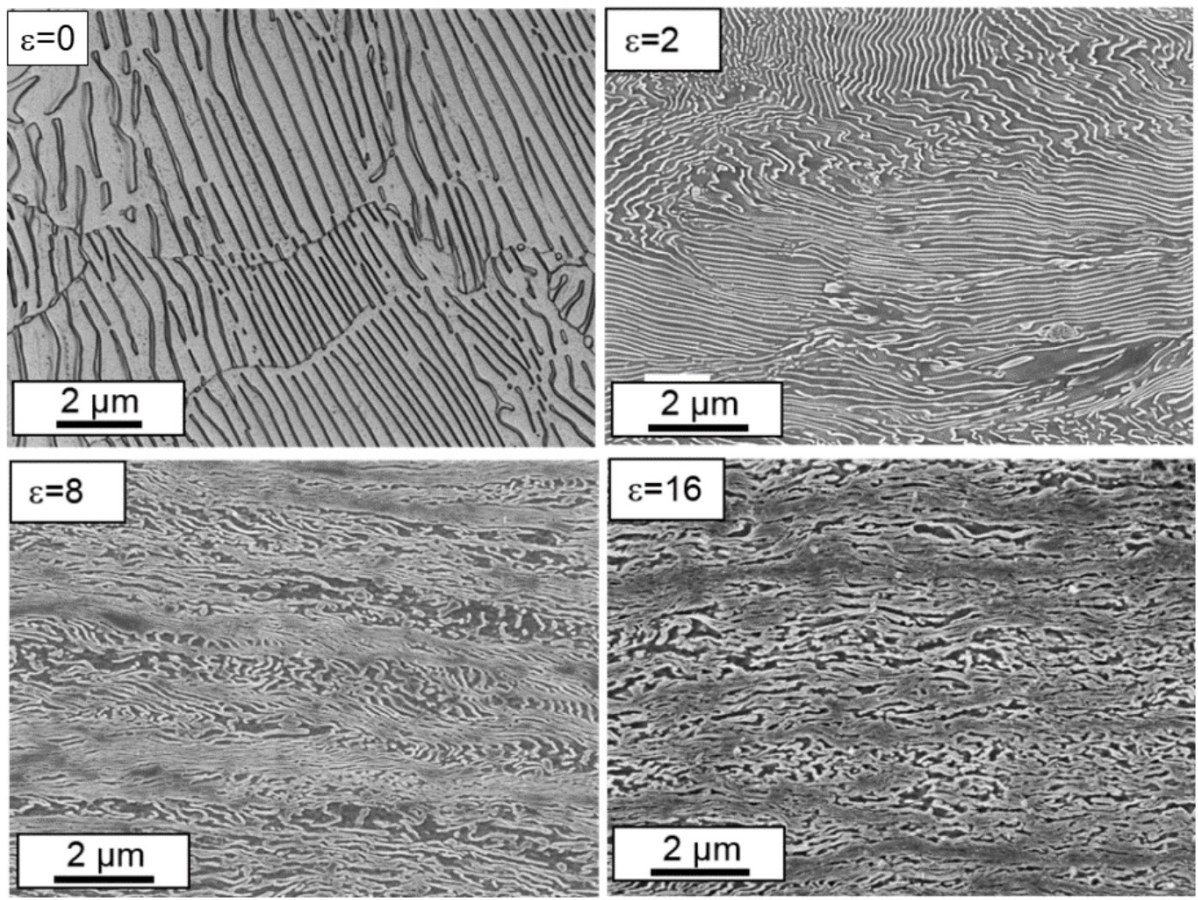

**Figure 13.** Scanning electron microscopy micrographs illustrating the structural evolution of the fully pearlitic rail steel R260 as a function of the equivalent strain (0, 2, 8 and 16) applied by HPT. Reproduced with permission[96].2010, Springer Nature.

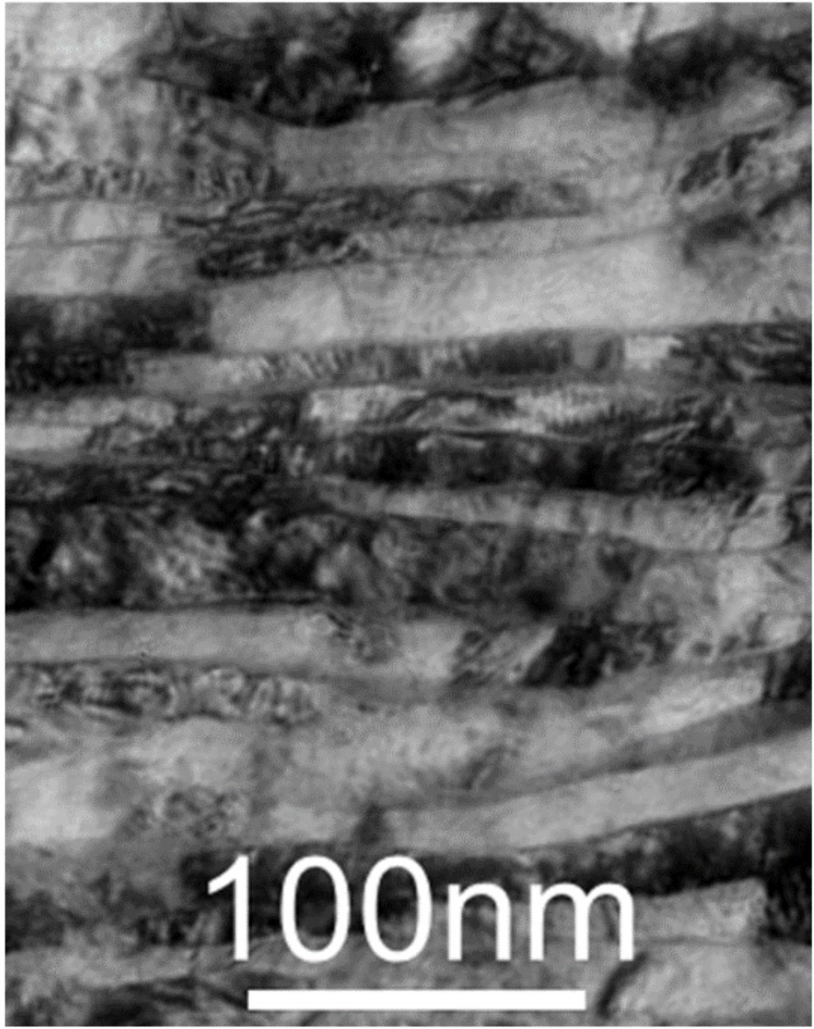

**Figure 14**. Transmission electron microscopy micrographs illustrating the structural of the pearlitic rail steel R260 at an equivalent strain 16 applied by HPT. Reproduced with permission[96].2010, Springer Nature.

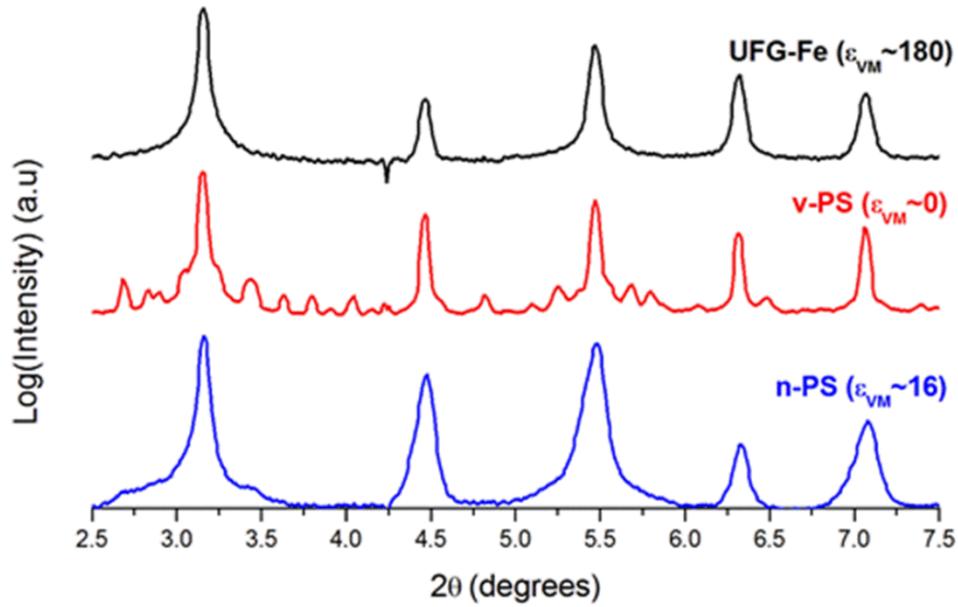

**Figure 15**. The integrated line profiles for ultrafine-grained iron (Fe) generated by HPT, an undeformed pearlitic steel R260 and a deformed full pearlitic steel R260 by HPT to a strain of about 16. Reproduced with permission[75].2018, Elsevier.

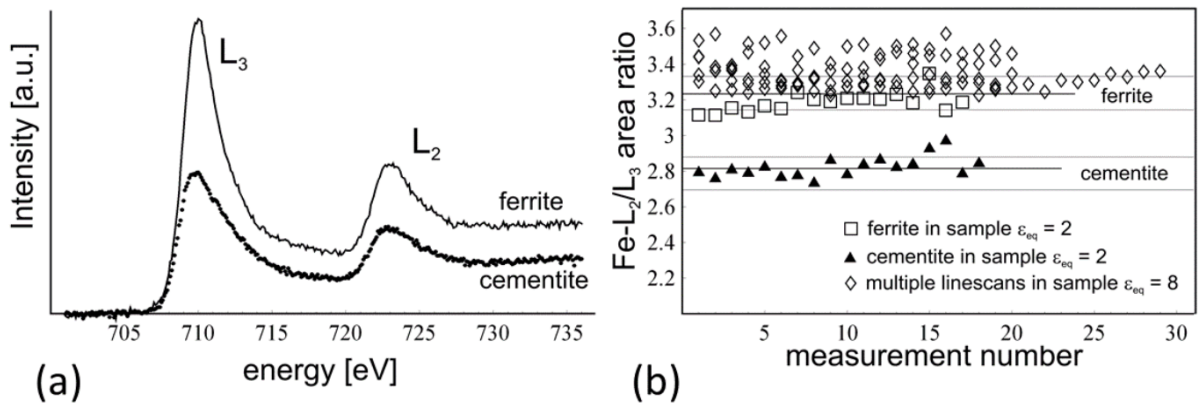

**Figure 16**. (a) Measured EEL spectra of ferrite and cementite of an undeformed full pearlitic rail steel R260 (a), Comparison of the area ratios of the EEL spectroscopy measurements of the starting material and samples with equivalent strain 2 and 8. The mean values for ferrite and cementite of the initial microstructure are indicated by black lines, the largest and smallest values are indicated by gray lines. Reproduced with permission[74].2006, Springer Nature.

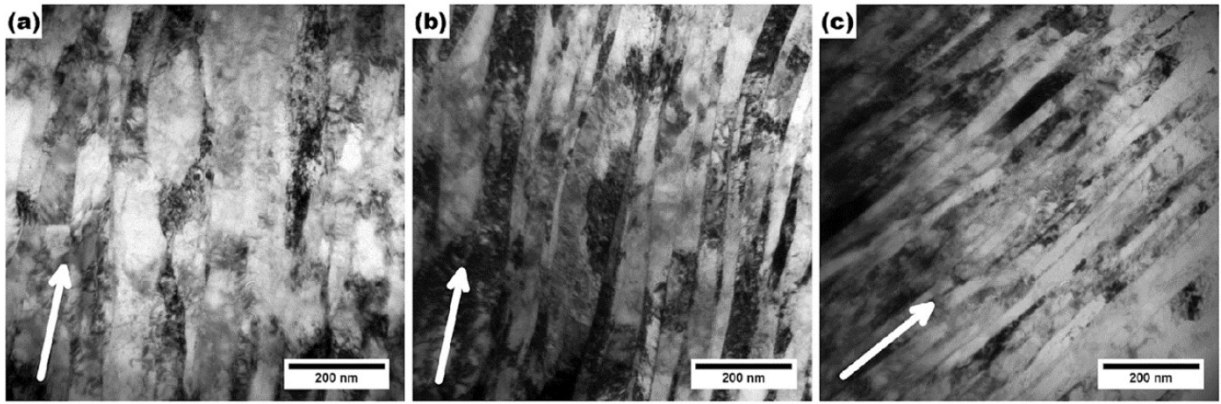

**Figure 17.** Transmission electron microscopy micrographs of a martensitic 0.1 wt.-% C steel after one turn HPT looking in radial direction. (a) and (b) are taken at a radius of r of 1mm (equivalent strain about 5) whereas (c) is from the same sample at r of 3mm (equivalent strain about15). The white arrows indicate the shear direction. Reproduced with permission[34].2019, Elsevier.

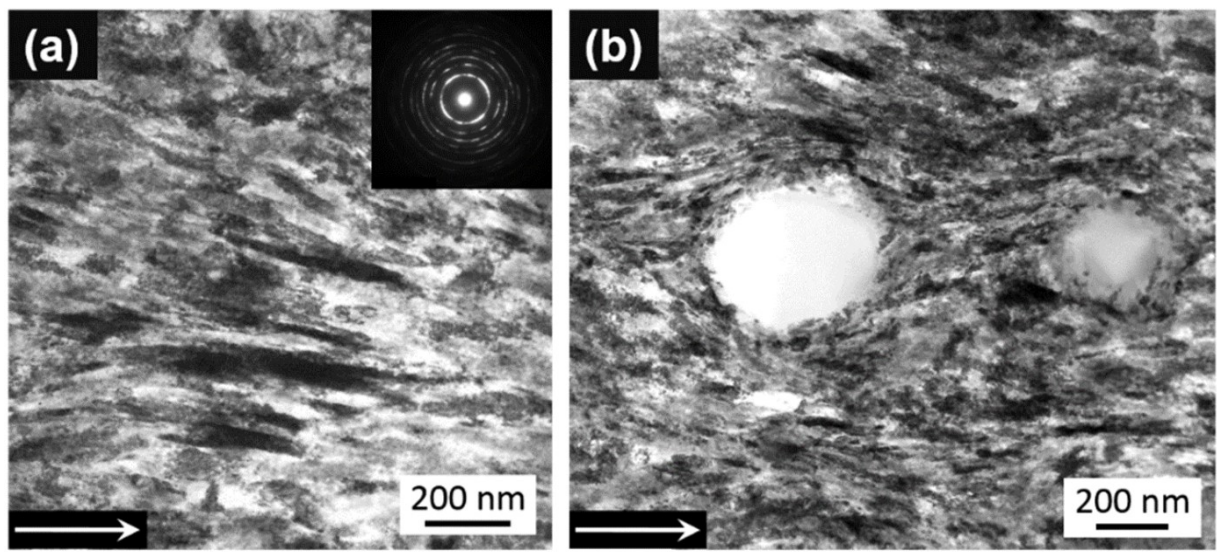

**Figure 18.** Transmission electron microscopy-micrographs showing the matrix microstructure of a HPT deformed 100Cr6 bearing steel, the applied shear strain was about 10. It exhibits that the evolved of the structure in the heavily deformed martensitic matrix is quite similar. (a) shows the morphology of the deformed matrix in a spheroidal cementite-free region and the corresponding SAED pattern; (b) indicate the flow of the deformed matrix around cementite precipitates. (a) and (b) are from the mildly deformed region. The white arrows mark the shear directions[92].

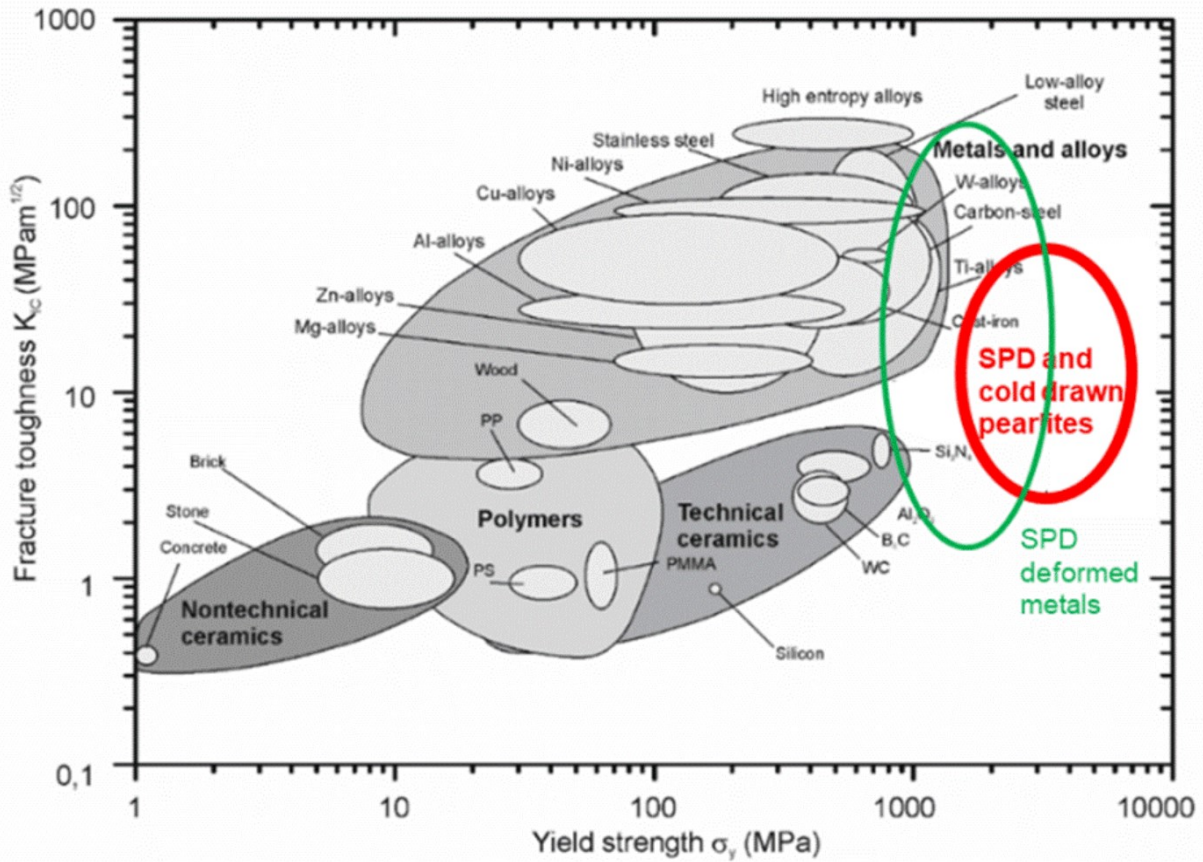

**Figure 19**. Ashby map illustrating the fracture toughness plotted against the yield strength of several classes of important engineering material. The SPD and cold drawn pearlitic steel variants set themselves apart from all other material classes[99] with strength values between 3–7 GPa and fracture toughness values between 4 and 40 MPa.m$^{1/2}$ depending on the testing direction. Even though they do not exceed the fracture toughness of several alloy groups, their damage tolerance in the main loading direction along the wire axis is exceptional and makes them actually to one of the most damage tolerant and strongest materials in the world. Reproduced under the terms of CC BY 4.0 licence[88]. 2016, Springer Nature.

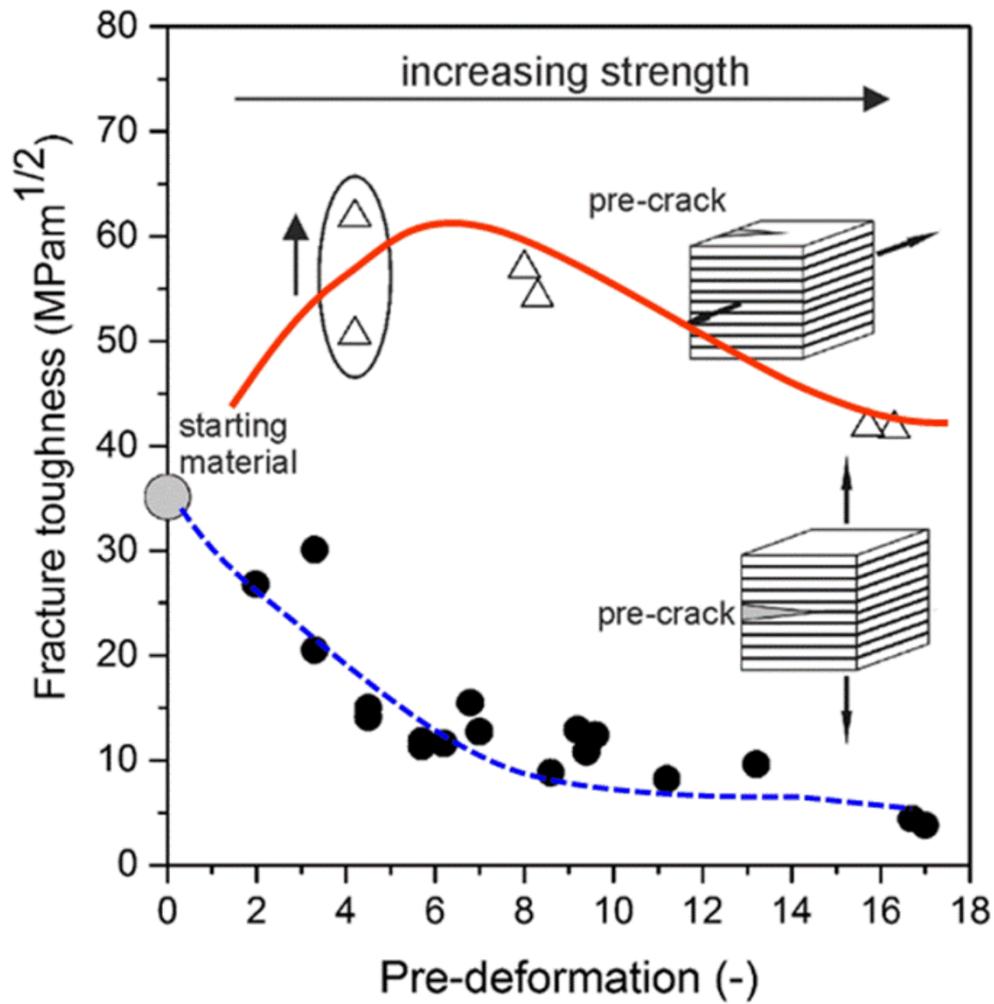

**Figure 20**. Fracture toughness for different crack plane orientations as a function of deformation strain by HPT of a full pearlitic rail steel R260. The crack plane in respect to the aligned lamella structure is indicated. Reproduced under the terms of CC BY 4.0 licence[29]. 2016, Taylor & Francis.

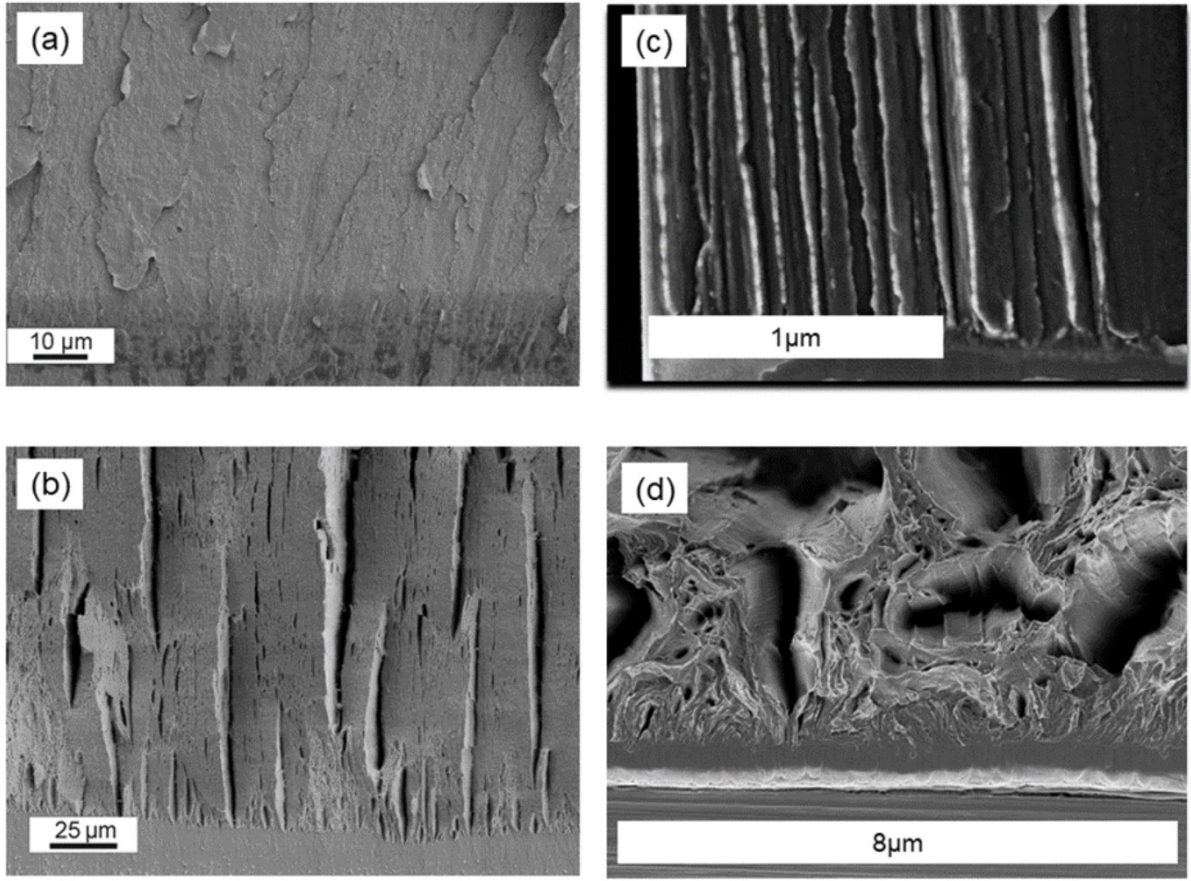

**Figure 21**. Scanning electron microscopy images of the fracture surfaces with the pre-crack parallel to the lamellae (a) for an HPT deformed R260 steel [96]and (c) the cold drawn pearlitic steel with the highest strength of 6.7GPa[88]. (b) and (d) shows the fracture surface where the pre-crack is perpendicular to the lamellae, (b) along the radial direction of the HPT disc of R260 steel with a strength of 3.5GPa and (d) perpendicular to the wire axis drawn with similar strength.   Reproduced with permission[96].2010, Springer Nature. Reproduced under the terms of CC BY 4.0 licence[88]. 2016, Springer Nature.

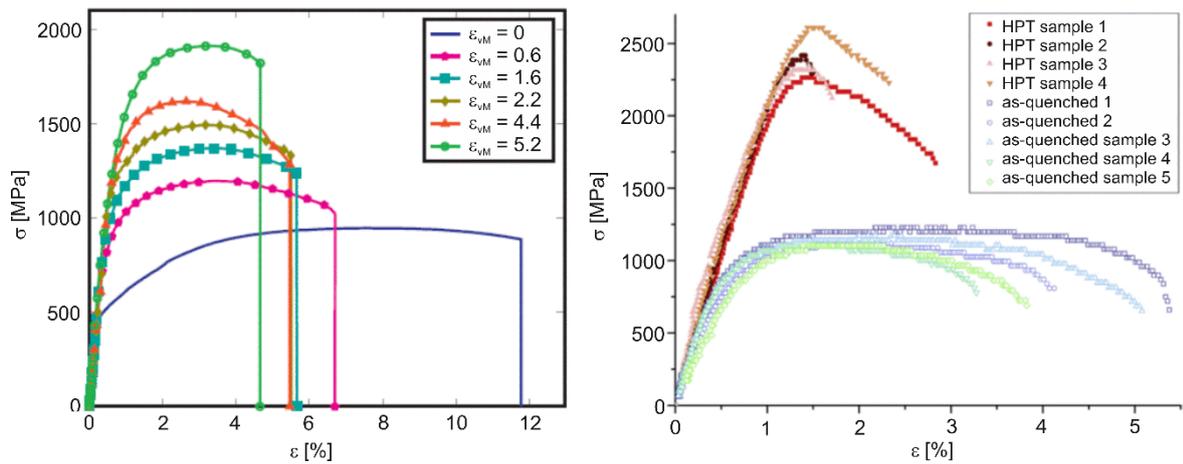

**Figure 22**. Engineering stress-strain curves of a full pearlitic rail steel R260 deformed to different strains by HPT [100] and a martensitic 0.1 wt.-% C steel as-quenched and HPT-treated (applied equivalent strain about 7.5) tested in tension along tangential direction[34]. Reproduced with permission[34][100].2015 and 2019, Elsevier.

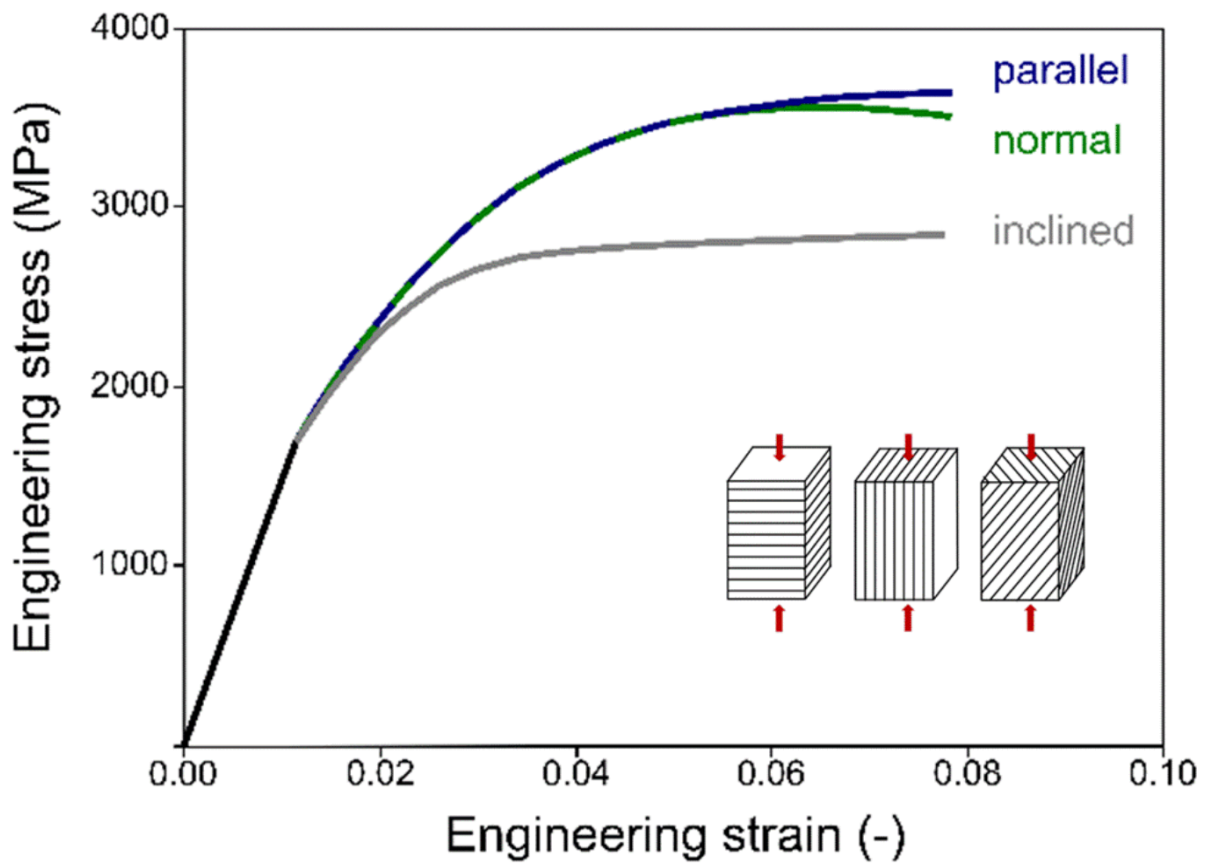

**Figure 23**. Illustration of the engineering stress-strain curves obtained in a micro compression test of a full pearlitic rail steel R260 deformed to an equivalent strain of 16 by HPT. Reproduced under the terms of CC BY-NC-ND 4.0 licence[73]. 2016, Elsevier.

**Table 1.** Hardness, *H*, grain size, *d*, and impurity concentration, *c*, of some pure nickel and nickel alloys subjected to HPT and a comparison to deposited NC specimens. The effectiveness of interstitial atoms on grain refinement becomes clearly visible.

| | c [at.-%] | synthesis | H [GPa] | grain size [nm] | Reference |
|---|---|---|---|---|---|
| Ni | < 0.01 | HPT at RT | 3.19 | 160 | [15] |
| Ni-Mo | 9.7 | HPT at RT | 5.24 | - | unpublished |
| Ni-Al | 9.3 | HPT at RT | 5.03 | 120 | [21] |
| Ni-C | 0.6 | HPT at RT | 5.15 | 90 | [15] |
| Ni-C | 3.5 | HPT at RT | 7.40 | - | unpublished |
| Ni (powder) | < 0.1 | HPT at RT | 5.25 | - | [22][23] |
| NC Ni | < 0.1 | electrodeposition | 4.72 | 21 | [24][25] |
| Ni-Mo-W | 14 (Mo) 2.4 | (W)DC sputtering | 8.95 | ~100[1] | [26] |

---

[1] note that within the grains planar defects (twins and stackig faults) with a 2 nm spacing are present.

**Andrea Bachmaier**

Andrea Bachmaier studied Materials Science at the University of Leoben and received her PhD degree in 2011. In 2013, she was awarded an Erwin-Schrödinger scholarship that allowed her to work on the decomposition process of supersaturated solid solutions and its influence on the thermal stability. In 2017, she received the noteworthy ERC Starting Grant. Currently, she is leading a research group at the Erich Schmid Institute of Materials Science, Austria. Her primary research focus is on the generation of metastable materials, novel nanocomposites and nanocrystalline metal-matrix composites by SPD and the investigation of their functional and mechanical properties.

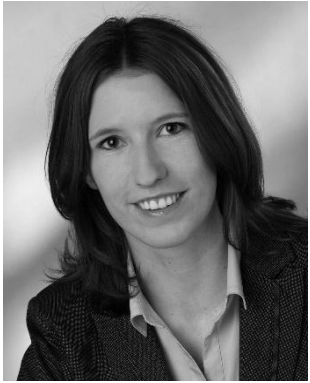

**Reinhard Pippan**

Reinhard Pippan graduated from the Technical University in Graz and then obtained his PhD degree at University of Leoben in 1982. His scientific career is mainly connected to the Erich Schmid Institute of Materials Science in Austria, where he remained for 38 years and where he was vice-director and group leader. Reinhard Pippan's research activities focus on the improvement of the basic understanding of the relations between the mechanical behaviour, the deformation processes, the fracture processes, and the micro- and nanostructure of the material. The effect of SPD on the structural evolution and the mechanical properties are a further research focus.

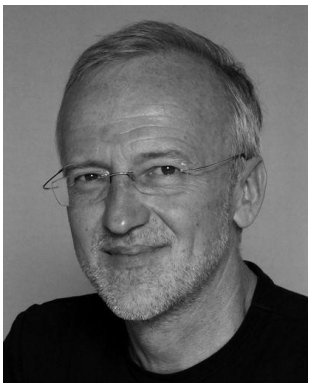

**Oliver Renk**

Oliver Renk studied Materials Science at the University of Leoben and received his PhD degree in 2015. He is currently a researcher at the Erich Schmid Institute of Materials Science, Austria. His primary research focus is on the structural evolution of metallic materials at large and severe strains and understanding the occurring migration processes of grain boundaries and how this grain boundary or interface plasticity could be used to improve and tailor mechanical properties of (nano)materials.

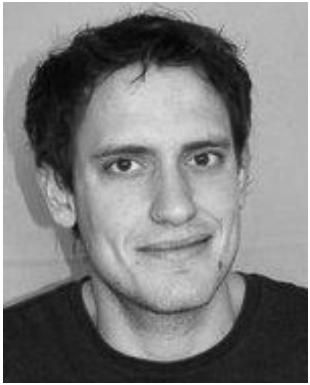

Andrea Bachmaier, Reinhard Pippan*, Oliver Renk

**Effect of carbon in severe plastically deformed metals**

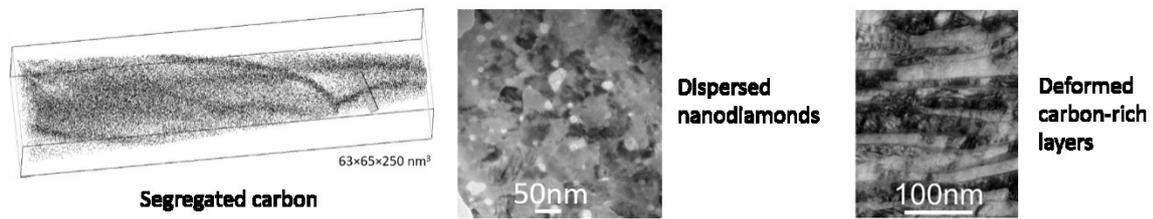

By adding carbon, the saturation grain size in severe plastically deformed materials can efficiently be reduced. Also other forms and allotropes of carbon (nanotubes, nanodiamonds or carbides) allow the generation of significantly finer microstructures. Independent of the type of carbon used high strength levels with exceptional ductility and toughness can be reached.